\documentclass[aps,pra,showpacs,amssymb,nofootinbib,twocolumn,longbibliography]{revtex4-1}
\pdfoutput=1
\usepackage[utf8]{inputenc}
\usepackage[english]{babel}

\usepackage{tikz}
\usetikzlibrary{shapes,backgrounds,fit,decorations.pathreplacing,arrows,decorations.markings}
\usepackage{verbatim}
\tikzstyle{vecArrow} = [thick, decoration={markings,mark=at position
   1 with {\arrow[semithick]{open triangle 60}}},
   double distance=1.4pt, shorten >= 5.5pt,
   preaction = {decorate},
   postaction = {draw,line width=1.4pt, white,shorten >= 4.5pt}]
\tikzstyle{innerWhite} = [semithick, white,line width=1.4pt, shorten >= 4.5pt]
\usepackage{bbm}
\usepackage{bm}
\usepackage{amsbsy}
\usepackage{amsthm}
\usepackage{amssymb}
\usepackage{amsfonts}
\usepackage{amsmath}
\usepackage{dsfont} 
\usepackage{graphicx} 
\usepackage{epsfig}
\usepackage{epstopdf}
\usepackage{dsfont}
\usepackage{multibib}
\usepackage{color}

\usepackage[colorlinks]{hyperref}
\makeatletter
\newcommand\org@hypertarget{}
\let\org@hypertarget\hypertarget
\renewcommand\hypertarget[2]{%
  \Hy@raisedlink{\org@hypertarget{#1}{}}#2%
  }
\makeatother
\usepackage[figure,table]{hypcap}
\usepackage{MnSymbol}
\usepackage{enumerate}
\usepackage{float}
\usepackage{csquotes}
\hypersetup{
	bookmarksnumbered,
	pdfstartview={FitH},
	citecolor={darkgreen},
	linkcolor={darkred},
	urlcolor={darkblue},
	pdfpagemode={UseOutlines}}
\definecolor{darkgreen}{RGB}{50,190,50}
\definecolor{darkblue}{RGB}{0,0,190}
\definecolor{darkred}{RGB}{238,0,0}
\definecolor{quantum}{RGB}{83,37,127}
\definecolor{quantumlight}{RGB}{169,146,191}
\usepackage{soul}
\newcommand{\djj}{d\kern-0.4em\char"16\kern-0.1em}

\renewcommand{\thesection}{\arabic{section}}
\renewcommand{\thesubsection}{\arabic{section}.\Alph{subsection}}
\makeatletter
\renewcommand{\p@subsection}{}
\renewcommand{\p@subsubsection}{}
\makeatother

\usepackage[customcolors]{hf-tikz}
\tikzset{style green/.style={
    set fill color=green!50!lime!60,
    set border color=white,
  },
  style cyan/.style={
    set fill color=cyan!90!blue!60,
    set border color=white,
  },
  style orange/.style={
    set fill color=orange!80!red!60,
    set border color=white,
  },
  style hordash/.style={
    set fill color=white,
    set border color=black,
  },
     style rose/.style={
    set fill color= magenta!70!pink!70, 
    set border color=white,
  },
  hor/.style={
    above left offset={-0.09,0.25},
    below right offset={0.09,-0.05},
    #1
  },
  ver/.style={
    above left offset={-0.09,0.35},
    below right offset={0.09,-0.1},
    #1
  }
}

\usepackage[framemethod=tikz]{mdframed}
\definecolor{mycolor}{rgb}{0.122, 0.435, 0.698}
\newmdenv[innerlinewidth=0.5pt, roundcorner=4pt,linecolor=mycolor,innerleftmargin=6pt,
innerrightmargin=6pt,innertopmargin=6pt,innerbottommargin=6pt]{mybox}
\usepackage{paracol}
\usepackage{tcolorbox}
\tcbuselibrary{theorems}

\newtcbtheorem{Definitions}{Definition}%
{colback=mycolor!5,colframe=mycolor,fonttitle=\bfseries,
left=4pt,
right=4pt,
top=5pt,
bottom=5pt}{defi}

\newtcbtheorem{Lemmas}{Lemma}%
{colback=green!5,colframe=green!50!blue,fonttitle=\bfseries,
left=4pt,
right=4pt,
top=5pt,
bottom=5pt
}{lemma}
\newtcbtheorem{Results}{Result}%
{colback=orange!5,colframe=orange!50!blue,fonttitle=\bfseries,
left=4pt,
right=4pt,
top=5pt,
bottom=5pt,
title={#2},
}{result}
\newtcolorbox[blend into=figures]{boxdefi}[3][]
{ float*=ht,width=\textwidth,lower separated=false, center upper,
title={#2},label= def:#3,#1}

\renewcommand{\thesection}{\Roman{section}}
\renewcommand{\thesubsection}{\Roman{section}.\arabic{subsection}}
\renewcommand{\thesubsubsection}{\Roman{section}.\arabic{subsection}.\arabic{subsubsection}}

\begin{document}

\title{“Philosophysics" at the University of Vienna:\\The (pre-)history of foundations of quantum physics in the Viennese cultural context}
\author{Flavio Del Santo}
\email{delsantoflavio@gmail.com}
\affiliation{Institute for Quantum Optics and Quantum Information - IQOQI Vienna, Austrian Academy of Sciences, Boltzmanngasse 3, 1090 Vienna, Austria}
\affiliation{Faculty of Physics, University of Vienna, Boltzmanngasse 5, 1090 Vienna, Austria}
\affiliation{Basic Research Community for Physics (BRCP)}
\author{Emanuel Schwarzhans}
\email{emanuel.schwarzhans@oeaw.ac.at}
\affiliation{Institute for Quantum Optics and Quantum Information - IQOQI Vienna, Austrian Academy of Sciences, Boltzmanngasse 3, 1090 Vienna, Austria}
\affiliation{Faculty of Physics, University of Vienna, Boltzmanngasse 5, 1090 Vienna, Austria}

\begin{abstract}
Vienna today is one of the capitals for the research on foundations of quantum physics. In this paper we reconstruct the main historical steps of the development of modern physics in Vienna, with an emphasis on quantum foundations. We show that the two main intuitive reasons, namely the influence of E. Schr\"odinger and the initiatives of A. Zeilinger in more recent years, cannot alone be held accountable for today's outstanding research landscape on foundation of quantum mechanics in Vienna. We instead show that the connection between physics and philosophy in Vienna always had an exceptional strength, and that this played a major role in establishing the prolific field of quantum foundations.  
\end{abstract}

\maketitle


\section{Introduction}\label{sec:introduction }

Vienna is today recognized  as a major center for the studies on the foundations of quantum theory (together with their modern applications to quantum information and quantum optics). This field of research is quite unconventional within physics, for it often addresses questions that lie at the border of philosophical investigation. Only in recent years it has become part of mainstream physics, largely by virtue of a plethora of practical applications, such as secure quantum communication and the promise of a universal quantum computer.

In this paper, our aim is to study the historical developments that made it possible for the foundations of quantum mechanics (FQM) to flourish in Vienna. More specifically, we will address the following question: \emph{was there something exceptional in the academic and, more generally, in the cultural landscape that made of Vienna a particularly fertile ground for the field of FQM to thrive compared to other places?}

At first sight, the answer seems quite straightforward: On the one hand, it is common knowledge that preeminent physicists concerned with FQM studied and worked in Vienna. Most notably, one of the founding fathers of quantum theory, Erwin Schr{\"o}dinger (1887-1961). Secondly, in the past few decades, the seminal initiatives of Anton Zeilinger (1945-) --one of the most influential living Austrian physicists-- have deeply shaped the research in physics at Austrian institutions.
Yet, this twofold explanation is not completely satisfactory: as we shall see, the influence of Schr{\"o}dinger cannot actually be held accountable, if not indirectly due to his renown, for the developments of the research on FQM in Vienna. Moreover, while it is unquestionable that Zeilinger --who indeed gave pivotal contributions to the field of FQM-- was the main driving force for the establishment of the research on FQM in Vienna, it is more that doubtful that this could have happened with the same success anywhere else. In fact, if we just distance ourselves from the “hagiographic" narrative which portraits great scientists as solitary heroes,\footnote{Hagiography is literally the idolized study of the lives of saints or religious leaders. This term is sometimes borrowed by historiography \cite{kragh} to indicate the widespread tendency of presenting the history of science as a succession of giants, who alone revolutionise their discipline.} often fighting against the \emph{Zeitgeist}, we should realize that the cultural context in which scientists live and work plays a fundamental role, not only for their intellectual formation, but also for the very development of novel ideas, and even more for the acceptance and the establishment thereof. In the specific case of Zeilinger, Freire's words reflect this idea:
\begin{quote}
Zeilinger’s intellectual style is marked by a deep curiosity, which was directed towards science during his undergraduate studies and favored by the flexible curriculum at University of Vienna at that time. In addition, he benefitted from Rauch’s support to research on foundations of quantum mechanics and from the intellectual climate of physics in Vienna --with its mix of science and philosophy-- a legacy coming from the late nineteenth and twentieth centuries  \cite{freire2014}.    
\end{quote}
Indeed, in the present paper --by means of original documents, as well as a collection of interviews with several physicists who have been protagonists (or close to them) in the development of modern FQM in Vienna--\footnote{All the interviews will be deposited in suitable institutional repositories, such as the archives of the Central Library for Physics of the University of Vienna, once they will be typewritten and edited.} we will try to retrace the main steps that helped  “loosen the soil" in which Zeilinger's school  eventually had the opportunity to plant the seeds that grew into today's Vienna excellence in FQM.
\begin{figure*}[ht!]
  \centering
  \includegraphics[width=\textwidth]{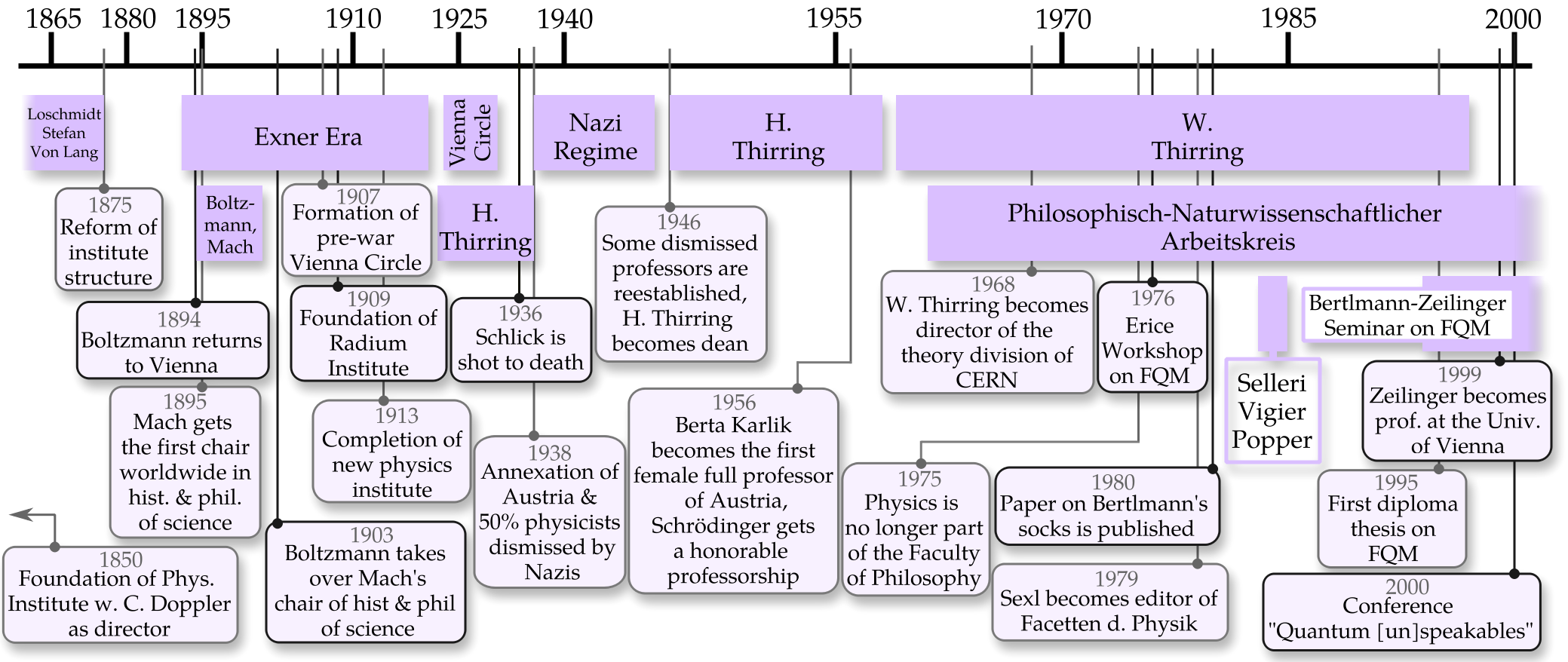}
  \vspace*{-1.5mm}
  \caption{{\footnotesize Timeline of selected events on matters of physics (and philosophy) in Vienna as recollected in the main text. The purple strips represents extended periods.}}
  \label{fig:chairs}
\end{figure*}

It is common knowledge that Vienna, at that time the capital of a multinational empire, became at turn of the twentieth century a major multicultural center for art, literature, philosophy, and science. This, as we shall see in what follows, reflected also in a flourishing landscape of intellectual debates at the edge between physics and philosophy, such as the Boltzmann-Mach controversy, or the Popper-Vienna Circle dispute about scientific method. This exceptional period of intellectual blossoming was shattered, like in most of Europe, by the advent of Nazi-fascism and the outburst of World War II. 

Before introducing our case study,  which will mainly focus on the research landscape in post-war Vienna, we ought to briefly recall the general situation of foundational research after World War II. In fact, the way of conducting science, and foremost physics, underwent a drastic change since the 1930s onward.  To begin with, the advent of Nazi-fascism (in particular the horrific enforcement of Nuremberg and Racial Laws in Germany and Italy, respectively) jeopardized the same community of physicists who had led the great revolutions of modern physics (the theory of relativity and quantum mechanics) in the previous couple of decades.

Moreover, during the war and immediately afterwards, with the transition into Cold War, a new paradigm took hold. A paradigm which saw the role of physicists changing dramatically: they became indispensable for the development of technologies, often of military nature (also following the success of the Manhattan Project in the US).\footnote{As a testimony of this, during a speech in 1951, a leading member of the US Atomic Energy Commission referred to physicists as a “war commodity”, a “tool of war”, and a “major war asset” to be “stockpiled” and “rationed” (quoted from \cite{kaiser}). 
 } This led to an unprecedented flow of funding into research in physics, which however, was coupled with a hyper-pragmatic and productivist attitude, mostly devoid of any fundamental and philosophical concerns. Such an attitude  went down in history with the emblematic expression “shut up and calculate!” (See e.g. \cite{kaiser}). Following the US --where this mostly originated-- as the western scientific leading force of the time, many European countries adhered to such an attitude and, as a matter of fact, research on quantum foundations virtually disappeared from scientific programs, with the only exception of a handful of “dissidents" (see Freire's comprehensive book \cite{freire2014}, and \cite{kaiser, baracca2017, besson2018, delsanto2019, freire2019, delsanto2020} for specific case studies). At the same time, even non-military research projects turned into a collective enterprise that involved tens if not hundreds or thousands of physicists (the so-called \emph{big science}) and it was predominantly oriented towards particle physics, characterised by the race to reach higher and higher energy scales in particle accelerators (such as the at that time newly founded CERN in Geneva). Also this kind of research was conducted pragmatically. Consequently, all foundational, philosophical or speculative aspects were regarded with suspicion or actively opposed.    
 
Also in Austria, and in particular in Vienna, a large part of the physics that was rebuilt in the post-war period focused on particle physics and was \emph{prima facie} conducted with the standard pragmatism of the time. However, as we shall see, what makes of Vienna an exceptional case is that there the drastic separation between science and its philosophical and fundamental aspects was never as complete as almost everywhere else. Firstly, at the institutional level, at the University of Vienna,  until as late as 1975 \cite{univie650} physics was a department of the Faculty of Philosophy. Therefore, the final exam (called \emph{Rigorosum}) to become \emph{Dr. Phil.} in physics included philosophy as a mandatory subject. Consequently,  Viennese physics students were receiving some formal education in philosophy. Secondly, although most of the teaching and research activities in Viennese universities were performed in a somewhat standard way as compared to other cities and countries, many leading physicists were actively involved in a number of initiatives that were rooted in a genuine interest towards philosophy of science and foundational research. Indeed, as Reinhold Bertlmann (1945\nobreakdash-) --an Austrian physicist who closely collaborated with John S. Bell (1928-1990) and, together with Zeilinger, was among the first ones to introduce FQM into teaching at the University of Vienna in the early 1990s-- nicely put it, 
\begin{quote}
officially they just followed this rule “shut up and calculate", but in their heart they were open and unofficially they discussed [foundational problems]; […] there was still an atmosphere of philosophy. \cite{bertlmanninterview}.
\end{quote}
It should be remarked that, contrarily to other cases (like Italy \cite{baracca2017}, or the US \cite{kaiser}) --where the interest towards foundations was revived between 1970s and 1980s, after a period of almost complete fading, and was motivated by political and ideological reasons-- in Vienna, philosophy and physics always maintained a relatively strong (although not always manifest) bond with a certain continuity. According to our account, this happened primarily for historical reasons, namely for the strength that this bond had acquired in \emph{fin-de-si\`ecle} Vienna. Indeed, Vienna has a reputation for its notorious cultural conservatorism, as crystallized by the famous \textit{dictum} attributed to the Austrian composer Gustav Mahler: “If the world ends, I would go to Vienna; there everything happens 50 years later".\footnote{Although we have no strong enough evidence to support the following additional thesis, we wish to remark that Austria never became a member of the North Atlantic Treaty Organization (NATO) --nor of the Warsaw Pact for that matter-- and in general the economical and cultural influence of US was there less pervasive than in other western European countries. This could have contributed to diminish the impact of military-oriented scientific research at Austrian institutions, and in general the significance of the “shut up and calculate" paradigm in Vienna.}

This tradition of philosophical interest and fundamental debate in physics in the Austrian capital was kept alive and recast by several physicists, such as theoreticians Herbert Pietschmann (1936-) and Roman U. Sexl (1939-1986), or the experimentalist Helmut Rauch (1939-2019). In particular, Pietschmann and the philosopher  Gerhard Schwarz (1937-) founded, in 1964, a “Philosophical-Scientific Working Group” (\emph{Philosophisch-Naturwissenschaftlicher Arbeitskreis}) \cite{philosophysik}, which catalysed the curiosity of physicists towards philosophical issues and allowed a structured forum of debate for physicists and philosophers.

Furthermore, as anticipated, despite one might think that the rise of foundational research could be greatly ascribed to the influence of Schr{\"o}dinger --who came back to Vienna in his late years (since 1956)-- we document that he deliberately did not establish a school, and did not essentially contribute further to FQM while in Vienna. Conversely, a series of external influences helped pave the way for the flourishing research environment on FQM in Vienna. This is the case of Franco Selleri (1936-2013) --the initiator of the revival of FQM in Italy \cite{baracca2017, delsanto2020}-- who had regular interactions with the Viennese community and even spent a sabbatical in Vienna where he lectured on FQM \cite{selleri1983}. Also, John S. Bell --who revolutionized the field of FQM more than anyone else in the post-war era, by formulating the inequalities that bear his name \cite{bell}--\footnote{Bell's theorem states that the  predictions of quantum mechanics are incompatible with those of any local hidden variable theory.} gave occasional talks in Vienna on the issue of local realism and he was the single major intellectual influence on Bertlmann.


\section{Physics (and philosophy) in Vienna before WWII}\label{sec:sub sec}

It is generally known that Vienna has an outstanding tradition of physics, which dates back to the second half of the nineteenth century, when the city experienced an unparalleled intellectual golden age (see, e.g., \cite{coen, schorske}). In those years, the University of Vienna saw among its professors physicists the likes of Christian Andreas Doppler (1803-1853), Joseph Stefan (1835–1893), Ernst Mach (1838-1916), Ludwig Boltzmann (1844-1906), and Franz Serafin Exner (1849-1926), who laid the groundwork for an exceptionally prolific physics milieu in the following years. The three Austrian Nobel Laureates in physics --Erwin Schr{\"o}dinger, Victor Francis Hess (1883-1964), and Wolfgang Pauli\footnote{Although Pauli never studied physics nor worked in any university in Vienna, arguably the intellectual Viennese landscape had a major influence on his career. Pauli's father was professor of Physical- and Bio-chemistry at the University of Vienna, whereas his godfather was Ernst Mach himself (Pauli's second name was Ernst after him). As a matter of fact, Pauli's knowledge of physics was already so advanced at the age of 18 that, before even leaving Vienna for starting his university studies in Munich, he already submitted a first original paper on general relativity \cite{pauli1919}. (See also the entry on Pauli's Viennese formative years of the IQOQI-Vienna blog “Bits of History": \href{https://www.iqoqi-vienna.at/blogs/blog/wolfgang-pauli-the-conscience-of-physics-came-from-vienna}{https://www.iqoqi-vienna.at/blogs/blog/wolfgang-pauli-the-conscience-of-physics-came-from-vienna})} (1900-1958)-- as well as Lise Meitner\footnote{The case of Lise Meitner --who completed her PhD in Vienna under Boltzmann and Exner in 1906-- is considered among the most unfair missed attribution of a Nobel Prize (she was nominated 48 times), because her long-term collaborator Otto Hahn was awarded the Nobel Prize in Chemistry in 1944, for their joint discovery of nuclear fission (see, e.g., \cite{meitner}).} (1878-1968) and Paul Ehrenfest\footnote{Ehrenfest, received his PhD in Vienna under Boltzmann in 1904. Although he never worked in Vienna, Boltzmann and his lectures had perhaps the deepest influence on Ehrefest scientific life \cite{ehrenfest}. Ehrenfest gave essential contribution to FQM, such as the theorem that bears his name.} (1880-1933) were all “academic children" of that extraordinary period.

It will be impossible to give a comprehensive account of the historical development of physics in Vienna in a short paper like this.\footnote {Although no such an account exists to date, we refer the reader to some of the relevant available literature (mostly in German): About the history of research and teaching at the University of Vienna, see \cite{univie650}. More specifically on the history of the main figures as well as of the Viennese Institutes of Physics, see the collection of short biographies of Austrian physicists published by the Austrian State Archives  \cite{biographies}, the Wiki edited by the City of Vienna \cite{physiker}, the booklet on the history of the Austrian Central Library of Physics \cite{bibliothek}, and a comprehensive book on Austrian female scientists \cite{wissenschafterinnen}. For an overview of the history of physics in Vienna before WWII and the historical scientific sites, see \cite{reiter}. Whereas, the most thorough piece of literature, unfortunately updated to 1949 only, is a doctoral thesis that reconstructs in detail the history of physics at the University of Vienna in the previous one hundred years \cite{bittner}. For an overview of the history of physics in Vienna in the 19th century see also \cite{histunivie} and references therein. See also the historical blog “Bits of History" of IQOQI-Vienna of the Austrian Academy of Science (\href{https://www.iqoqi-vienna.at/blogs/bits-of-history}{https://www.iqoqi-vienna.at/blogs/blog/bits-of-history}), and the book on the history of the research on nuclear physics in Austria \cite{kerne}.}
However, with the hope of making this paper as self-contained as possible, in this section we will provide an overview of some of the central intellectual figures and the main historical happenings on matters of physics until War World II, with a focus on the interplay between physics and philosophy. The reader already acquainted with such topics, can directly jump to Section \ref{postwar}, where the exposition of the main original part of this research begins.

\subsection{Towards modern physics}
The era of “modern" physics in Vienna began in 1850 with the foundation of the new \emph{Physikalisches Institut} (“Physical Institute") and the appointment of Christian  Doppler  as professor of Experimental Physics and director of the Institute of Physics.  This was part of a general renovation of Austrian universities implemented by the Minister Leo Graf Thun-Hohenstein as a consequence of the March Revolution of 1848, which had started within the University of Vienna. Due to these renovations, the University moved “in buildings scattered through the outer districts" \cite{schorske}, and the new Institute of Physics was located far away from the city centre in the suburbs of Vienna-Erdberg (Landstraße 104, today Erdbergstrasse 15), where student movements could not so easily organise protests \cite{univie650}. From 1863, Josef Stefan joined the Institute (of which he became director in 1866), followed by the young lecturers Victor von Lang (1838-1921), Ludwig Boltzmann and Ernst Mach. Josef Loschmidt (1821-1895) was also appointed associate professor in Vienna in 1868 and full professor of Physical Chemistry in 1871 \cite{histunivie}.
The years until 1875 were particularly prolific: Stefan, together with his student Boltzmann, became renowned for the Stefan-Boltzmann law relating the power radiated by a black body to the fourth power of its temperature, which allowed to estimate the surface temperature of the sun; moreover he was among the first physicists to popularize Maxwell's seminal works in continental Europe. von Lang was among the pioneers of crystal physics and the first to establish this field in Austria \cite{biographies}. Mach, still a student at that time, was the experimentally demonstrated the Doppler effect for the dependence of the sound-wave frequency on the movement of its source \cite{reichenbach}.\footnote{The Doppler effect had been already experimentally demonstrated by the Dutch chemist Christophorus H. D. Buys Ballot in 1945. However, Mach's experiment put an end to a lasting debate on the interpretation of the Doppler effect, in favour of Doppler and against the alternative view of the Viennese mathematician Joseph Petzval \cite{reichenbach}.} Loschmidt contributed to several fields of physics and chemistry, notably to the kinetic theory of gases, where he is remembered for the “paradox" that bears his name \cite{wu}.

\subsection{Physics meets philosophy: The age of Mach and Boltzmann}
\label{machbol}
 In 1875, the Institute of Physics moved back closer to the city centre, in T\"urkenstrasse 3, for a “provisional arrangement” (which ended up lasting until 1913) and was divided into the “Physical Cabinet" (from 1902 “1st Physics Institute", concerned with experimental physics), the “Institute of Physics" (from 1902, “Institute for Theoretical Physics") and the “Physical-Chemical Institute" (from 1902, “2nd Physics Institute") \cite{bittner}, see Fig. \ref{fig:chairs}. In 1891, Franz Exner took over the chair of Loschmidt, whereas Boltzmann returned to Vienna in 1894 as Stefan's successor. 
 
  At that time, Boltzmann was already renowned for his pioneering work in one of the most prolific fields of modern physics, statistical mechanics, which led him to be “generally acknowledged as one of the most important physicists of the nineteenth century" \cite{uffink}. Remarkable are his kinetic theory of gases and the fundamental formula $S=k_B W$, relating the entropy $S$ of a system with the probability of occurrence of its macroscopic state $W$ through the constant $k_B$ named after him. Moreover, in 1872, Boltzmann had proven the H-theorem, which provided the statistical explanation of the second law of thermodynamics.\footnote{For a comprehensive account of Boltzmann's contribution in statistical physics, see e.g. \cite{uffink, boltzmannbio}, and  \cite{boltzmannpapers} for an edited collection of all his works.} The main criticism to Boltzmann's H-theorem came from Vienna in 1877, when Loschmidt  pointed out that no time-asymmetry can arise purely from (Newtonian) mechanics and would need some additional assumptions (in 1896 Ernst Zermelo put forward a similar argument based on Poincaré recurrence theorem; see, e.g., \cite{wu, uffink}, for the development of this debate in irreversibility).
  
Since 1892 --which, incidentally, roughly coincides with his return to Vienna-- Boltzmann started devoting part of his intellectual activity to philosophy, a tendency that “mirror[ed] increasing doubts about the validity of Newtonian mechanics, his own work on kinetic theory, and even the atomic theory." \cite{blackmore1999}. Also due to his vast and faceted intellectual production, still today there is no consensus on what Boltzmann's \emph{Weltaschauung} was (or even if there was a consistent single view), but it is undoubtful that Boltzmann's philosophy has been influential and explicitly praised by many (especially Viennese) physicists with philosophical proclivities (e.g. Schr\"odinger) and philosophers proper (such as Karl Popper and Paul Feyerabend)  \cite{blackmore2013}.
Regardless of the complexity of his philosophical views (on physics and its philosophy), the fact that Boltzmann's scientific work hinted at the reality of (at that time unobservable) atoms and molecules not only informed his philosophy, but became to be known as its main feature. This  has later been identified with a form of materialism.\footnote{Even Vladimir Lenin, in 1909, praised Boltzmann's philosophy as aligned with materialism \cite{lenin}.} 

This view was antagonised by Mach, who by that time had become one of the most influential intellectuals in Vienna (see \cite{blackmore2001}). In 1897, for instance, Boltzmann gave a talk at the Academy of Sciences in Vienna, and Mach, who was among the audience, commented: “I don't believe that atoms exist!"  \cite{bachtold}.\footnote{Note that the narrative of a fierce personal fight between Mach and Boltzmann, which, according to certain accounts, could have even been among the causes that led Boltzmann to commit suicide, has been greatly scaled down in the recent historigraphical literature \cite{stadler2019, blackmore2013}.}  Indeed, Mach's philosophy was characterised by an anti-metaphysical (thus anti-realist) attitude to which he opposed a strictly empiricist approach, according to which science should aim at describing natural phenomena in the most economical way (Mach's ideas have been a major influence on, among others, the development of Einstein's relativity theory and on the views of the Vienna Circle, see Sect \ref{circles}). Despite his background in physics, in 1895 Mach fully transgressed the disciplinary boundaries by being appointed to what could be regarded as the first chair worldwide in history and philosophy of science, “in particular the history and theory of inductive sciences”, created for him at the University of Vienna \cite{stadler2019}. In 1903, after Mach's retirement (due to a stroke in 1901), that chair was to be occupied by Boltzmann, until he took his own life in 1906.

This intellectual debate between Mach's empiricism and Boltzmann's realism had a lasting impact to foundations of physics, general philosophy of science, and it gave momentum to the interplay between physics and philosophy. This was even more so in Vienna, about which Schr\"odinger recalled the intellectual environment of his formative years:
\begin{quote}
I was born and educated in Vienna with E. Mach's teaching and personality still pervading the atmosphere. [...] Just as strong or even stronger than Mach's was in this time in Vienna the after-effect of the great Boltzmann [...]. Their views were not the same. But filled with a great admiration of the candid and incorruptible struggle for truth in both of them, we did not consider them irreconcilable.\footnote{Letter from Schr\"odinger to Sir Arthur Eddington on March 22, 1940. (Archive of the Österreichische Zentralbibliothek für Physik: \href{http://phaidra.univie.ac.at/o:260913}{http://phaidra.univie.ac.at/o:260913}).}
\end{quote}

\subsection{Intellectual Circles}
\label{circles}

In the midst of this cultural and intellectual blossoming of Vienna, the beginning of the twentieth century saw the foundation of two outstanding circles of intellectuals concerned with physics and philosophy: The “Exner Circle" and the “Vienna Circle".

\subsubsection{The “Exner era"}

Around the turn of the twentieth century, Franz Serafin Exner became one of the most influential 
physicists in Vienna. His political, didactic and scientific endeavours, besides his own and his disciples' scientific achievements, were to leave a deep and lasting mark on the scientific landscape of Vienna. For this reason some refer to this period as the “Exner era" \cite{histunivie}. In the following, we will elucidate Exner's role and the influence of his school on the scientific culture in Vienna and at the international level, and  how this 
laid the foundations for basic research for generations of Viennese scientists to come. \\

Exner grew up in an academic family. His father, Franz Serafin Exner (Father) (1802-1853) was an influential philosopher and reformer of education, which most likely had had an influence on his son's philosophical and cultural views. The latter studied physics at the University of Vienna with his most renown teachers being Josef Stefan, Viktor von Lang and Johann Josef Loschmidt.\footnote{Loschmidt's talent was discovered early on by Franz Exner (Father) who consequently supported and mentored the young destitute student (\cite{karlik1982}, p. 16 and 33).} In 1891, Exner got a full professorship, first for the the Physical-Chemical Institute of the University of Vienna, previously led by  Loschmidt \cite{karlik1982}.

Amongst his students and peers, Exner was highly valued for his teaching skills, but even more important was his social engagement, which was partly the reason why his institute developed into the social center of physics research in Vienna, despite the desolate equipment of the university facilities at that time.\footnote{Exner was of the opinion that Science is only a part of human culture, and that the arts should be on equal footing with science \cite{karlik1982} p. 61.} 
 This environment attracted a large number of students, many of whom eventually became  highly influential scientists. The group that formed around Exner this way was later called the “Exner Circle" (\textit{Exner-Kreis}) \cite{karlik1982}: Among them  were two Nobel laureates, Viktor Franz Hess and Erwin Schrödinger, as well as Lise Meitner,\footnote{It was only since 1897 that in Austria was allowed for women to attend lectures at the faculty of philosophy, therefore in physics. The first woman who completed a PhD (with a major) in physics was Olga Steindler in 1903. Lise Meitner was the second in 1906. Both graduated under Exner and Boltzmann.} Viktor Conrad, Felix Eherenhaft, Felix Maria Exner-Ewarten, Friedrich Hasen\"ohrl, Stefan Meyer, Egon Schweidler and Hans Thirring. 
 Due to his engagement with his students, Exner has been referred to as “father of the next generation of Vienna physicists" \cite{reiter}. His role as such is exemplified in a letter from Erwin Schrödinger on the occasion of the recent death of Exner, in which he thanked Stefan Meyer 
\begin{quote}

    for sending the commemoration of our deceased father [Exner], because that is what he really was to us [...] \cite{karlik1982} 
\end{quote}

Exner, a full-fledged experimentalist, was interested in a wide spectrum of research areas. His main contributions in physics can be collected into four research fields: electro-chemistry, atmospheric electricity, spectral analysis and colour theory. Apart from his research contributions, Exner's  social and organizational engagement put Vienna in the international spotlight. In fact, Exner's involvement in the foundation of radioactivity research in Austria is of particular merit. Amongst many connections he established with physicists all over the world,\footnote{In 1883, Exner became editor of the scientific journal \textit{Carls Repertorium der Physik}, which he led for about 20 years. This position put him into contact with highly esteemed physicists all over the world. For example the work of Michelson and Morley, about the relative motion of the Earth with respect the \textit{luminiferous aether}, was publish there.} 
his involvement in the promotion of the research of Marie Sklowdowska Curie (1867-1934) and Pierre Curie (1859-1906) is of particular interest.

Upon the request of Marie and Pierre Curie to the Austrian government to relinquish some pitch blend\footnote{Also called uraninite, a raw material that contains uranium and radium.} from its mining facilities, the government requested advice from the Austrian Academy of Sciences. It was Exner who advised the government to support the Curies,\footnote{Due to his friendship with Wilhlem C. R\"ontgen, with whom he had worked in his time in W\"urzburg, Exner was aware of the potential of radioactivity research. In fact, already in 1896, a few weeks after Röntgen had discovered X-Ray radiation, Exner reported possible medical uses of this new radiation to the society of doctors in Vienna \cite{karlik1982}.} who thus extracted radium for the first time. In appreciation of his recommendation, in 1899, Exner received a small sample of highly radium enriched pitch blend extract from the Curies. This enabled him and his students to conduct research on radioactivity in Austria. Due to Exner's ongoing interest and strong involvement in the research on radioactivity, the Austrian Academy of Science appointed him as chair for the newly founded commission for radioactive substances. In 1909, partly as a consequence of Exner's efforts\footnote{The institute was built upon the initiative of Karl Kupelwieser (1841-1925), who donated 500,000 Austrian kronen for its construction \cite{karlik1982}.}, the Austrian Academy of Sciences started building the Institute for Radium Research,\footnote{Exner was entrusted with planning the construction of the institute, which he did with the help of Stefan Meyer, in Boltzmanngass 3 (today the same  building hosts the IQOQI-Vienna).} the world's first institute fully devoted to the research on radioactivity (\cite{karlik1982}, pp. 93).\footnote{The Radium Institute was not only special for being one of the first of its kind, but for its particular gender distribution of researchers. Most remarkably, between 1919 and 1934, one third of the active researchers were women who published as much as man did. For such a gender diverse environment to be possible at that time it required a particular institutional and political context, fostered by a unique constellation of progressive politics and supportive, politically aware personalities. See Rentetzi's entry of the IQOQI-Vienna blog “Bits of History": \href{https://www.iqoqi-vienna.at/blogs/blog/why-women-scientists-thrived-at-the-radium-institute-in-interwar-vienna}{https://www.iqoqi-vienna.at/blogs/blog/why-women-scientists-thrived-at-the-radium-institute-in-interwar-vienna}, and references therein.} Exner was instated as the head of this institute, alongside with Stefan Meyer as acting director.\footnote{Stefan Meyer became the head  of the institute in 1920 until the \textit{Anschluss} in 1938 when, as a Jew, he was forced to leave office (see section \ref{anschluss}).} 

Moreover, Exner was entrusted with planning the construction and the furnishing of the new building for the Physics Institutes, located in the Boltzmanngasse (Weisenhausgasse until 1913), corner Strudlhofgasse, next door to the Radium Institute. When the constructions was ultimated in 1913, all three Physics Institutes moved to this building complex, which still hosts the Faculty of Physics today. It is worth mentioning that in the same year of the inauguration of the new building, an outstanding physics congress --both for the prominence of the participants and for its size-- took place in Vienna:  
 Over 7000 physicists gathered to discuss on the state of the art of modern physics, including Einstein's theory of general relativity and radioactivity. The speakers included Einstein himself, who gave a lecture on “The present status of the problem of gravitation" which stimulated in many physicists the interest toward  general relativity, including Schr\"odinger  \cite{moore}, p. 70.


Of particular interest are also Exner's epistemological views, which undoubtedly had an influence on his disciples, in particular on Schr\"odinger. Already in 1907, when he became the Rector of the University of Vienna, in his inauguration speech Exner made the following, at that time quite controversial statement, wherein he hinted at the fundamentally indeterministic nature of physical processes:
\begin{quote}

    Everything that happens in nature is the result of random events. [...]  
     If the probability that the same regularity emerges every time is so great that it becomes a certainty for human concepts, then we speak of a law. This however, is only possible with such a large number of events beyond all conceptions, such as occur in molecular processes. In all other areas there are no laws, only regularities, and these become all the more doubtful the smaller the number of events from which they are derived, and finally, when the number of them is too small, turn into random phenomena.
\end{quote}
Upon his return to Vienna, Schr\"odinger points out in his book  “Was ist ein Naturgesetz" \cite{schr_nat}  that Exner had already, seven years before quantum mechanics was developed, conceived of the idea of irreducible indeterminism \cite{karlik1982}\cite{flavioindet}.\footnote{In the philosophical literature this view has been called “Vienna indeterminism" \cite{viennaindet}. This view has in fact been conceived by both Exner and Boltzmann. Schrödinger also emphasizes that the development of quantum mechanics has brought Exner's circle of ideas, “by the way, without Exner's name ever being mentioned", into the focus of interest (\cite{karlik1982}, pp 82).}

As we mentioned before, the Exner Circle gathered several well known and highly important scientists who continued shaping the scientific landscape worldwide. However, of particular interest to the developments of physics in Vienna are Felix Ehrenhaft, Friedrich Hasen\"ohrl and Hans Thirring. 

Ehrenhaft finished his studies at the University of Vienna in 1903. 
In 1908 he began investigating the elementary electricity quantum (nowadays called the electron). Due to his discovery of the standard method of measuring the electric charge and his subsequent experiments he acquired fame throughout Europe. However, soon afterwards these developments, his reputation was strongly damaged due to several controversial reports he made (e.g. measurement reports contradicting the assumed charge quantisation). When Exner retired as a professor in 1920, Ehrenhaft was endorsed by some of his colleagues as a successor of Exner to the lead of the “Second Physical Institute", yet he did not become director. In fact, his --at that point questionable reputation as well as his high degree of specialization (as opposed to Exner, who valued many facets of cultural and scientific life)-- disqualified him for the position\footnote{ The university of Vienna consulted Einstein for this decision, who was against the indication of Ehrenhaft.}. Nevertheless,  the University of Vienna  opened a new institute for him, called the 3rd Physics Institute, partly because he was seen as a valuable experimentalist by some of his peers \cite{karlik1982}, but, more importantly, for what appears to have been political decisions to mitigate his influence \cite{santos, Braunbeck2003}. 

 Friedrich “Fritz" Hasen\"ohrl completed his doctoral dissertation under Exner in 1897, and was appointed the chair of Theoretical Physics previously held by Boltzmann, after the latter's death. It should be mentioned here, that Hasen\"ohrl anticipated Einstein's famous energy-mass equivalence already in 1904 (in a special case, for cavity radiation)  \cite{Hasenohrl1904-2, Hasenohrl1904-3}.\footnote{This won him an invitation to the prestigious first Solvay Conference. See Del Santo's entry on Hasen\"ohrl in the IQOQI-Vienna blog “Bits of History": \href{https://www.iqoqi-vienna.at/blogs/blog/a-forgotten-viennese-physicist-friedrich-hasenoehrl}{https://www.iqoqi-vienna.at/blogs/blog/a-forgotten-viennese-physicist-friedrich-hasenoehrl}, and references thereof.} Hasen\"ohrl was seen as a fine lecturer and  his course of theoretical physics was up to date with the state of the art: He taught special relativity, since around as early as 1912, and the first quantum effects --Planck's quantisation of energy and the photoelectric effect-- already in 1911 \cite{kuhn}. It  was, to a high degree thanks to this course that Schr\"odinger and Hans Thirring dedicated their lives to theoretical physics \cite{moore, kuhn}.\footnote{Schr\"odinger went so far as stating: “No other person had a stronger influence on me then Fritz Hasen\"ohrl, except perhaps my father." \cite{moore}.} Hasen\"ohrl prematurely died as a volunteer in World War I. 
 
Hans Thirring wrote his dissertation under Friedrich Hasen\"ohrl and, in 1910, he became his assistant.\footnote{Schr\"odinger, who had actually been the first choice, was at that time doing his military service. He then became assistant to Exner in his experimental group.}
In  1918 --together with his colleague, the Viennese physicists Josef Lense (1890-1985)-- Thirring published his first influential paper on what became to be known as the Lens-Thirring effect  \cite{lense}.\footnote{Which introduces a correction to the precession of a body spinning in a gravitational field, due to general relativistic effects. It has been claimed that Einstein, with whom Thirring was in regular contact, had a major role in the discovery of this effect (see \cite{pfister}).} In 1921 Hans Thirring became assistant professor and director of the Institute for Theoretical Physics, and, finally, full professor in 1927.\footnote{Worth mentioning is that among Thirring's students was Victor Weisskopf (who studied at the University of Vienna between 1926 and 1928), who then --under Thirring's recommendation-- moved to G\"ottingen to complete his doctoral studies with Max Born.} For what concerns quantum physics, as early as 1929, Thirring published --together with his former student and assistant Otto Halpern-- one of the first books which gave a systematic account of the newly developed quantum theory \cite{halpern1929}.\footnote{Indeed, this book, reprinted in English translation in 1932,  was one of the first ones to present both Heisenberg's matrix mechanics and Schr\"odinger's wave mechanics approaches, as well as de Broglie's matter waves. To the best of our knowledge only a book by George Birtwistle --who was lecturer in Cambridge-- was published before Halpern and Thirring's book, in 1928, with the same features \cite{birtwistle1928}.} Remarkably, the last chapter of the book is fully devoted to the “interpretations of the theory", thus showing a high consideration for foundational issues. 

\begin{figure}[ht!]
  \centering
  \includegraphics[width=\linewidth]{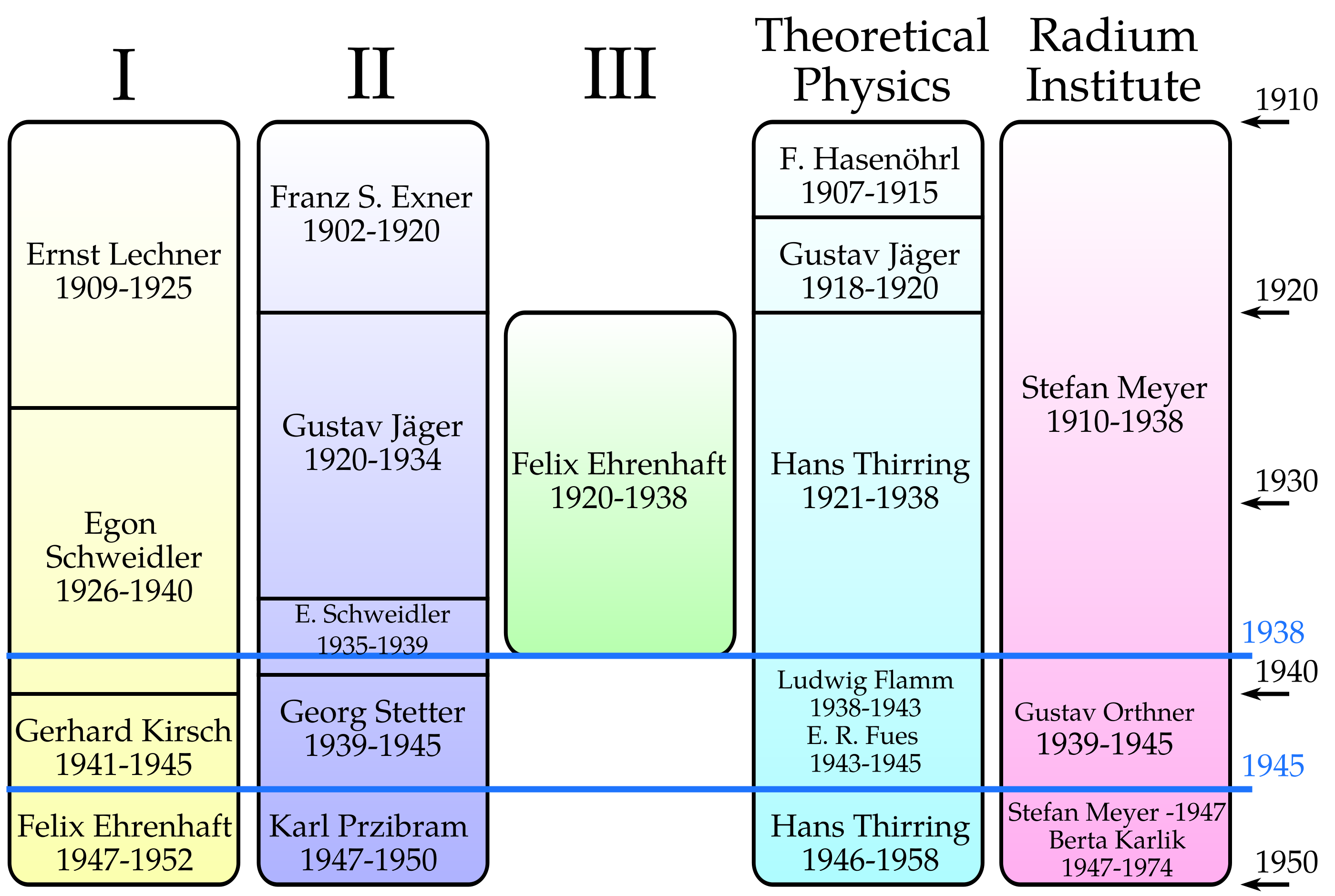}
  \vspace*{-1.5mm}
  \caption{{\footnotesize Graphic depicting the directorships of the different physics institutes in the years between 1910 and 1950. The roman numbers stand for the first, second and third Institutes of Physics. The blue lines mark the annexation of Austria to Nazi Germany with the subsequent replacement of the professors (1938), and the the end of the second world war, after which some professors were reinstated. We thank Franz Sachslehner for providing us with the table that this graphic is a modified version of: [\href{https://phaidra.univie.ac.at/o:1208930}{https://phaidra.univie.ac.at/o:1208930}, under the CC BY 4.0 license] }}
  \label{fig:chairs}
\end{figure}

\subsubsection{The Vienna Circle}
Overlapping with the time when the “Exner circle" was active, around 1907, another group of students started regular discussion meetings --they would meet on Thursday evenings in a Viennese coffeehouse-- on the fundamental problems of science and philosophy. The main members of this group were Philipp Frank (1884-1966) --who studied physics, mathematics and philosophy, and who wrote his dissertation under the supervision of Ludwig Boltzmann--,\footnote{Which, however, he  was only able to finish after Boltzmann's death.} Hans Hahn (1879-1934), a student of mathematics who was later to become known for the Hahn-Banach theorem, and Otto Neurath (1882-1945), who studied mathematics, economy and history.\footnote{Among other things, Neurath became famous for his method of pictorial statistics, the International System of Typographic Picture Education (ISOTYPE) method. At the time of the “Red Vienna" (1919-1934), Neurath also became a member of the Social Democrats.}
Strongly influenced by philosophical questions connected to the works of Bertrand Russel, Max Planck, David Hilbert and Albert Einstein, as well the legacy of Mach and Boltzmann, their discussion usually gravitated around one central topic, as Frank writes half a century later:
\begin{quote}
How can we avoid the traditional ambiguity and incomprehensibility of philosophy? How can we bring philosophy and physics closer together again? \cite{sigmund2015}
\end{quote}
However, this pre-war discussion circle dissolved not much later and would reunite only after the First World War to form the group that later became known as the \emph{Vienna Circle}.

In the early 1920s, after Hans Hahn had moved back to Vienna and Otto Neurath had been forced to move back to Austria,\footnote{After trying to implement his idea of “Vollsozialisierung", \emph{i.e,} “complete socialisation" in a politically highly unstable Bavaria, in the aftermath of several political coups he got arrested and later deported to Austria.} Moritz Schlick (1882-1936), a philosopher who was known for his work on the philosophy of Einstein's theory of relativity, was appointed professor at the University of Vienna, taking the chair for natural philosophy.\footnote{In a way inheriting the chair of Ernst Mach and Boltzmann; see section \ref{machbol}.} Urged by his students, in particular Friedrich Weissmann (1896-1959) and Herbert Feigl (1902-1988), in 1924, Schlick (together with Hans Hahn and Otto Neurath) founded a discussion group initially called the “Schlick Circle", which, however later became to be known as the “Wiener Kreis" (\emph{Vienna circle}).\footnote{The discussion group officially called themselves Vienna Circle for the first time in their manifest \emph{Die wissenschaftliche Weltauffassung}  published in 1929. The name \emph{Vienna circle} was an idea of Otto Neurath as a response to a, in his opinion, rather dry title. It should have reminded the readers of pleasant things (such as “Wiener Walzer" and “Wienerwald").}

Further regular members that also were part of the pre-war discussion circle were  Philipp Frank, the mathematician and philosopher Olga Hahn (1882-1937) --Hans Hahn's sister-- and the philosopher Viktor Kraft (1880-1975).\footnote{Viktor Kraft was one of the few members of the Vienna Circle that experienced and survived the Second World War in Vienna. During the Nazi regime he lost his position at the University. However, after the war, in 1950 he became full Professor of Philosophy at the University of Vienna. It is worth mentioning here that Paul Feyerabend (1924-1994), one of the most influential philosophers of science of the twentieth century, graduated in philosophy under Viktor Kraft in 1951. Feyerabend had also attended courses of Felix Ehrenhaft, Hans Thirring and Karl Przibram, members and descendants of the Exner circle, before moving from physics to philosophy. See Collodel's entry in the IQOQI-Vienna blog “Bits of History": \href{https://www.iqoqi-vienna.at/blogs/blog/paul-feyerabend-between-physics-and-philosophy}{https://www.iqoqi-vienna.at/blogs/blog/paul-feyerabend-between-physics-and-philosophy}, and references thereof. After WWII, Feyerabend founded in Vienna a philosophical group of students around the figure of Kraft, which revived the interests of the Vienna Circle (empiricism, anti-metaphysics, etc.), called  “Kraft's circle" or the “Third Vienna Circle" \cite{thirdviennacircle}.} 
Additionally, young bright minds such as the philosopher Rudolf Carnap (1981-1970), the mathematician Karl Menger (1902-1985), the logician Kurt Gödel (1906-1978), as well as Edgar Zilsel (1891-1944), Felix Kaufmann (1895-1949) and Gustav Bergmann (1906-1987) became regular members of the discussion circle. Besides the above mentioned influences by  Planck, Hilbert, Einstein and Russel on the Vienna Circle, Ludwig Wittgenstein's book \emph{Tractacus logico-philosophicus} (firstly published in 1918) strongly shaped the early views of the circle. Although he was repeatedly asked to, Wittgenstein never subscribed as a member of the circle, nor did he join their meetings. In contrast to Wittgenstein, Karl Popper (1902-1994) would have felt it an honour to be invited by the Vienna Circle, but never was.\footnote{As Popper recalls in \cite{popper1992}: “Feigl writes [...] that both Edgar Zilsel  and I tried to preserve our independence 'by remaining outside the Circle'. But the fact is that I should have felt greatly honoured had I been invited, and it would never have occurred to me that membership  in  Schlick’s  seminar  could  endanger  my  independence  in  the slightest degree."}  Nevertheless, Popper was in regular contact with members of the Vienna circle, in particular with ``secondary circles" lead by Viktor Kraft, Edgar Zilsel and Karl Menger, who were interested in Popper's views. The latter, however, developed in those years his harsh criticism of Vienna Circle's \textit{logical positivism}, a position that regards induction and verification at the foundations of science, putting forward in 1934 his most famous alternative to the problem of induction and demarcation based on \textit{falsifiability} \cite{popper34}.
 
 From 1924 to 1933, the Vienna Circle met every Thursday in a small seminar room at the ground floor of the physics building of the University of Vienna at Boltzmangasse 5.
 Like its precursor before the war, the Vienna Circle was mostly concerned with the interplay of science
 and philosophy. 
It became one of the most central actors for the development of the philosophical movement of logical empiricism.\footnote{Along side with the "Berlin circle" and other individuals that were in contact with either of these two groups.} Its members strongly criticised metaphysical considerations and identified apparent philosophical statements as pseudo-statements if their content could not be subjected to empirical scrutiny. They advocated the viewpoint that 
``everything that lies beyond the factual must be viewed as meaningless." \cite{passmore1967}.
 In particular this viewpoint lead to a partial reintegration of science into philosophy, as it became widely agreed ``that philosophy should make use of technical devices, derived from logic, in order to solve problems relating to the philosophy of science, that philosophy is not about `the world' but about the language through which men speak about the world" \cite{passmore1967}.

\subsection{The end of the ``golden age"}
\label{anschluss}
Beginning with the 1930s, the Vienna Circle started to dissolve due to the increasingly hostile political environment which led to the rise, in 1933-34, of the Austro-fascist regime of Engelbert Dollfuß and later Karl Schuschnigg. The 
Vienna Circle came to a violent end in 1936, when Moritz Schlick was shot to death by his former student Johann Nelböck.\footnote{Nelböck had a long-lasting personal feud with Schlick. He stood in strong opposition to Schlicks philosophical views and, already when he was a student of Schlick, he showed up in front of Schlick's office, threatening to kill him. Nelböck got arrested immedeately after killing Schlick, however, in 1938 he successfully requested a pardon from the German authorities, arguing that the murder was a act of ``ideological and political necessity"\cite{sigmund2015}.} After Schlick's death some of the members of the Vienna Circle continued meeting in a sporadic fashion. The annexation of Austria to Nazi Germany in 1938 (\textit{Anschluss}) and the subsequent ``cleansing" of the University for racial and political reasons, however, put an end to the Vienna Circle, and to the exceptional intellectual climate of Vienna as a whole. 

In physics as well the situation rapidly degenerated as some of the most distinguished figures of the Viennese landscape were forced to leave office in 1938.\footnote{For an overview on the exile of Austrian scientists after the \textit{Anschluss}  see \cite{reiter88,graf, medawar,feichtinger2014}.} Sharing the same regrettable fate of an enormous number of European intellectuals, around 50\% of Viennese physicists were dismissed, and 40\% of them forced to emigrate \cite{reiter88}. Notably, Stefan Meyer, Felix Ehrenhaft, as well as the renown Austrian experimental physicists, Karl Przibram (1878-1973) --a pupil of the Nobel Laureate Joseph J. Thomson \cite{przibram}-- were expelled from Austrian institutions because they were Jews. They had to leave the country (with the exception of Meyer who due to his political acquaintances was granted the right to retire into private life in the countryside), and, in the case of Ehrenhaft, ``he was arrested by the police and beaten up, had his money confiscated", and the ``3rd Physics Institute" of which he was the head was dismissed (see also footnote \ref{denazi}) \cite{santos}.  Also Hans Thirring was sent to forced retirement because of his ``corroding influence on the military readiness of the nation", and his friendship with Jewish intellectuals the likes of Einstein and Sigmund Freud \cite{feyerabend}.\footnote{Interestingly, after Thirring's forced retirement, the chair of theoretical physics remained vacant until 1943, and one of the candidates for that position was Werner Heisenberg (1901-1976), who expressed his intention to take that appointment (see \cite{kerne}, p. 255). However, due to not fully clarified circumstances, Heisenberg did not become professor in Vienna, and that chair was instead appointed to Ludwig Flamm (1885-1964), who was already professor at the \textit{Technische  Hochschule} of Vienna.}

 In conclusion, the unparalleled intellectual ``golden age" that had characterized Vienna in the late 19th and early 20th centuries was abruptly shattered by the annexation of Austria to Nazi Germany. We will see that, especially in the case of physics, one would have to wait another two decades for a proper reconstruction to take place. Moreover, after WWII, the academic culture underwent a substantial change: the long-standing symbiosis between physics and philosophy that had characterized a great deal of the European research environment --with Vienna at the forefront-- came to an end. As we shall see, although also in Vienna this separation definitely took place, the deep intertwinement between physics and philosophy kept smouldering under the surface.

\section{The post-war era}
\label{postwar}

\subsection{Reconstructing physics in Vienna: Walter Thirring}\label{wthirring}

Like most of the other activities, reconstructing science in Austria after the Nazi period and the War required a great effort, and quite some time (it should also be recalled that Austria got back its state sovereignty only in 1955). For what concerns physics, Hans Thirring was reestablished as the only professor of Theoretical Physics at the University of Vienna, Przibram as head of the  “2nd Physics Institute" \cite{przibram}, and also Ehrenhaft came back to Vienna from the US, becoming the director of the “1st Physics Institute".\footnote{\label{denazi}It should be remarked that the \emph{denazification} and the restoration of the status quo at Austrian institutions was very mild (see also \href{https://geschichte.univie.ac.at/en/articles/de-nazification-professorate-university-vienna}{https://geschichte.univie.ac.at/en/articles/de-nazification-professorate-university-vienna}). In physics, Ehrenhaft --whose “3rd Physics Institute" which he founded and directed before the War was dismissed by Nazis and never restored--  returned to Vienna in 1947 only with the status of visiting professor and for this reason, despite the despicable political circumstances of his forced flee, Austria did not even grant him retirement benefits \cite{thirring}. Moreover, quite disgracefully, after Ehrenhaft's death, his successor to the directorship of the “1st Physics Institute" was Georg Stetter (1895-1988), who had been previously dismissed from the University of Vienna after WWII because he was a Nazi and a top member of the German nuclear energy project \cite{biographies}.} 

Although Thirring became the director of the Theoretical Institute and the dean of the Faculty for Philosophy in 1946, inevitably his intellectual priorities had somewhat changed. Immediately after the War, indeed, he started devoting a great deal of his interests to pacifistic and political initiatives. Paul Feyerabend, who was a student of physics in Vienna in those years, recalled him saying: “This is important, physics is not" \cite{feyerabend}. In particular, Thirring became a pioneer of anti-nuclear movements, writing a book on the “history of the atomic bomb" \cite{atombombe}, and being the only Austrian representative in the famous Pugwash Conferences.\footnote{The Pugwash Conferences on Science and World Affairs is an association founded in 1957 in the course of a conference in Pugwash, Nova Scotia --to which Thirring participated-- following the initiative of the Russel-Einstein Manifesto against nuclear weapons. It was awarded the 1995 Nobel Peace Prize for its achievements towards denuclearization. See also Broda's short article “Pugwash and Austria" (Archive of the Österreichische Zentralbibliothek für Physik: \href{https://phaidra.univie.ac.at/o:922132}{https://phaidra.univie.ac.at/o:922132}).} Some years later, Thirring also engaged in professional politics, serving as a member of the Austrian Federal Council (\emph{Bundesrat}) for the Socialist Party of Austria (today Social Democratic Party of Austria, SP\"O) between 1957 and 1964 \cite{biographies}. In that role, he promoted further pacifistic initiatives, such as a proposed unilateral disarmament of Austria, known as the “Thirring-Plan".

Thirrings deep involvement in such complex political challenges, however, most likely came at the expense of, both, his own scientific achievements and the quality of his physics teaching. Potentially, this was one of the reasons why university courses in Vienna lagged behind the actual scientific developments at that time. Pietschmann, who had started his studies in physics at the University of Vienna in 1954, recalls that “it was kind of transition period. In those days there was no quantum physics taught at the University of Vienna.[...] Not at all!" \cite{pieinter}.\footnote{According to Pietschmann, Hans Thirring “did not know anything about quantum physics" \cite{pieinter}. Clearly this cannot be the case since Thirring had authored one of the first books in this field (see section \ref{circles}) and he had been supervising theses on quantum (wave) mechanics, such as the one of Paul Urban (1905-1995) in 1935 --who was to become a well-known expert in that field \cite{biographies}. The fact that Thirring did not teach quantum physics seems to corroborate the thesis that his commitment had largely shifted away from physics after the War.} The only other theoretician --a lecturer (\emph{Dozent}) who was said to know quantum physics-- was Theodor Sexl (1899-1967), who, however, also never taught it in his courses. A story circulating among the students at that time tells that he had written a book on quantum mechanics which was lost in the mail on its way to the publisher, therefore he refused to teach quantum physics.\footnote{Peter Aichelburg (1941-), who started studying physics in Vienna in 1959, remembers that Theodore Sexl gave a “course on theoretical physics but this was purely classical [...]. [It] was a bit disappointing because it seemed very old fashioned." \cite{aichelburg}. And Pietschmann recalls that Sexl taught Theoretical Physics for two semesters and instead of quantum mechanics, he would teach the outmoded Bohr-Sommefeld atomic model until the 1960s.}   Whether there is some truth in this anecdote or otherwise, as a matter of fact, after WWII neither Hans Thirring nor Theodor Sexl, nor any other professor for that matter, taught quantum physics at the University of Vienna until as late as 1959! \cite{thirring}.\footnote{Pietschmann recently recalled that a lecturer (\textit{Dozent}) shortly taught quantum physics at the University of Vienna at some point between the end of the War and 1956, but he left for US (see Pietschmann's entry in the IQOQI-Vienna blog “Bits of History": \href{https://www.iqoqi-vienna.at/blogs/blog/hans-thirring-a-personal-recollection}{https://www.iqoqi-vienna.at/blogs/blog/hans-thirring-a-personal-recollection}).} Remarkably, the students showed a great deal of initiative and used to meet in self-organised groups after the official lectures to study quantum physics  on the famous book \textit{Principles of Quantum Mechanics} by Paul Dirac \cite{dirac}.\footnote{Pietschmann, who with his fellow students experienced this, also points out that in those times the so-called “mass university" was not yet born and the students of theoretical physics were altogether twelve \cite{pieinter}.}

In fact, it was only from the mid-1950s that the real post-war modernization of Viennese physics took hold. Firstly, Berta Karlik (1904-1990) was appointed professor of Experimental Nuclear Physics in 1956, being the first woman in the whole history of Austria   to become a full professor (\emph{ordentliche Professorin}) \cite{biographies}.\footnote{In 1973, Karlik also was to become the first female full member (\textit{wirkliches Mitglied}) of the Austrian Academy of Sciences.} Moreover, for what concerns fundamental physics, at the beginning of the same year, the Nobel laureate Erwin Schr\"odinger returned to Vienna as “honorary professor" (\emph{Ehrenprofessor}), where he remained until his death in 1961.\footnote{His return was arranged by Hans Thirring, who informed Schr\"odinger of his appointment in Vienna in a letter on January 23, 1956. (Archive of the Österreichische Zentralbibliothek für Physik: \href{https://phaidra.univie.ac.at/o:129804}{https://phaidra.univie.ac.at/o:129804}).} This surely changed the landscape of Viennese physics, causing great excitement among students, who, however, were to see some of their expectations betrayed. It is true, on the one hand, that Schr\"odinger stimulated the young physicists through his though-provoking talks, in which foundational problems played a central role, such as his inaugural address on “The Crisis of the Atomic Concept" \cite{moore}, and that he engaged in public philosophical debates, for instance when Victor Weisskopf (1908-2002) visited Vienna in 1958 \cite{moore, pieinter}. Yet, his impact on the Viennese research and education for what concerns foundational issues, especially those dealing with quantum mechanics, turns out to be much more limited than one can think. 

First of all, during his years as a professor at the University of Vienna, Schr\"odinger mostly taught the course ``General Relativity and Expanding Universes" (his very last lecture was held in March 1958) \cite{moore}, and only once around 1956 he gave a lecture on “wave mechanics": “Of course not on quantum mechanics, but on wave mechanics; Schr\"odinger never used the words quantum mechanics", Pietschmann --who attended some of these courses-- recollects \cite{pieinter}. Indeed, Schr\"odinger remained coherent throughout his career to what he once had said to Niels Bohr (1885-1962) many years before: “If we are still going to have to put up with these damn quantum jumps, I am sorry that I ever had anything to do with quantum theory” \cite{moore}. This attitude had even more dramatic consequences, because the awareness of being a dissident about his views on quantum theory persuaded Schr\"odinger to purposely not take any PhD students and thus prevented him to form a school, at least in Vienna. An anecdote told by Pietschmann, who directly experienced such a refusal, best explains this:
\begin{quote}
[Schr\"odinger] said he could not take any graduate student because he was working only in two fields: one was general relativity and he simply had no good ideas. And then, you know, his eyes lightened with sparks coming out almost, and he said: “the other is wave mechanics. There, of course --he said-- I have lots of ideas”. And then he said, and I kept this in mind: “I cannot take the responsibility to put a young man on a track which is considered to be a dead end by the rest of the world.” \cite{pieinter}.\footnote{See also Pietschmann's entry in the IQOQI-Vienna blog “Bits of History": \href{https://www.iqoqi-vienna.at/blogs/blog/erwin-schroedinger-professor-at-the-university-of-vienna}{https://www.iqoqi-vienna.at/blogs/blog/erwin-schroedinger-professor-at-the-university-of-vienna}.}
\end{quote}

On the contrary,  perhaps the most remarkable change for what concerns the modernization of physics in post-war Vienna  stemmed from the retirement of Hans Thirring and the stir caused by the search of a new professor of theoretical physics. According to Schr\"odinger's biographer, “the various factions were agreed only on one thing: to prevent Schr\"odinger from using his enormous prestige to influence the decision." \cite{moore}. In the end, quite shockingly, it was decided to appoint as the successor of Hans Thirring, his son Walter (1927-2014). In Pietschmann's words:
\begin{quote}
there is one kind of nepotism which is worldwide unique, and this is Walter Thirring. I mean, no doubts about his capabilities, but he inherited the chair from his father. So Schr\"odinger was against it [...] because he said it is simply impossible to inherit a chair at the University. \cite{pieinter} 
\end{quote}

 Walter Thirring studied physics at the Universities of Innsbruck and Vienna, where he graduated under Felix
Ehrenhaft in 1949.\footnote{Walter Thirring's first and life-long passion was music, which he cultivated by becoming an organ player and composer. It was only after the untimely death of his older brother Harald during WWII that Walter Thirring was called to "carry on the physics tradition in the family." \cite{thirring}.} After his PhD, most likely also thanks to the influential acquaintances of his father, the young Thirring fulfilled the dream of any physicist: he went on a “Grand Tour"  to perfect his formation under some of the most distinguished living physicists of that time (and fathers of quantum theory). In 1949, he was in Dublin with Schr\"odinger, and the following year in Glasgow with Bruno Touschek (1921-1978). Between 1950 and 1952, he was firstly in G\"ottingen with Werner Heisenberg and then in Zurich with Wolfgang Pauli. He spent the years 1953-1954 in the US at Institute for Advanced Study in Princeton, where he had contacts with Einstein, before becoming lecturer at the University of Bern (Switzerland) for two years, and finally visiting professor at the  Massachusetts Institute of Technology (Boston, US), where he met Victor Weisskopf, until 1958.\footnote{For a complete account of this period, see Thirring's intellectual autobiography \cite{thirring}.} During these remarkable years, Thirring gave outstanding contribution to the development of the quantum theory of fields and in particular to quantum electrodynamics, about which he published his first book as early as in 1955 \cite{qed}. In 1958, his renown suddenly increased when he formulated the first non-trivial exactly solvable interacting model of a field theory (for Dirac fermions in 1+1 dimensions), the \textit{Thirring model}.\footnote{For a commented collection of the main scientific results of Walter Thirring, see \cite{thirworks}.} So, when Walter Thirring finally came back to Vienna in 1959 as a full professor, he was an internationally recognised expert in quantum field theory.

Not only was Thirring to become perhaps the most influential Austrian physicist in the following decades, but he was also the real initiator of post-war physics in Vienna. 
First of all, he was the first professor to finally teach quantum mechanics, together with more advanced courses on contemporary physics. He also played a major role in making Austria become a member state of CERN --where he was to be the director of its Theory Division between 1968 and 1971-- that helped put Austria back on the map for what concerns physics research at the international level. Moreover, Thirring attracted many young students who completed their dissertations with him, both on (theoretical) high energy physics and on general relativity.  Many of that new generation of students (and the immediately following one) were to become professors at the University of Vienna and rebuilt the physics landscape there. In particular, two of Thirring's first students in Vienna, Herbert Pietschmann and Roman Ulrich Sexl (the nephew of the physicist Theodor Sexl) became professors at the end of the 1960s,\footnote{Many students of Thirring who gathered in the group of Sexl became professors of relativity and gravitational physics at the University of Vienna, such as Peter Aichelburg, Helmut Urbantke, Helmut Rumpf, and Franz Embacher.} and, as we shall see in the next section, helped restore the Viennese tradition of connecting physics with philosophy.

\subsection{Reconnecting physics with philosphy: Pietschmann and Sexl}\label{sec:sub sec}

\subsubsection{The \textit{Philosophisch-Naturwissenschaftlicher Arbeitskreis}}
\label{PNA}

As we have seen, Walter Thirring quickly became a renowned  mathematical physicist and, under his lead, theoretical physics in Vienna experienced a new period of prosperity. However, Thirring himself did not show a particular interest towards philosophical or interpretational issues, favouring formal problems of theoretical high energy physics: “He was very pragmatic”, recalls Peter Aichelburg, a former student of his who was to become professor of Gravitational Physics in Vienna \cite{aichelburg}.\footnote{Pietschmann also points out that “[Walter] Thirring didn't know anything about philosophy." \cite{pieinter}. Only at the very end of his career, in fact, Thirring devoted some of his work to Bell's inequalities, and even entered some debate on philosophy of science in connection to religion.} Nevertheless, Thirring was not only very tolerant but, to a certain extent, even supportive of the initiatives that some of his collaborators and former students, foremost Pietschmann and Sexl, started organising in the mid-1960s. 

Subsequently to the refusal by part of Schr\"odinger to take him (or anyone else, for that matter) as a student, Pietschmann completed his PhD at the University of Vienna under Walter Thirring in 1960 and, after some international experiences at University of Virginia (US) and at the University of Bonn (Germany), he came back to Vienna as a professor in 1968. Only a few months later, however, Walter Thirring accepted the aforementioned directorship of the Theory Division of CERN, and Pietschmann remained the only professor of Theoretical Physics in Vienna --Theodor Sexl (who actually never became full professor) had died the year before-- and \textit{de facto} became the head of the Theoretical Physical Institute. Pietschmann, contrarily to his mentor Thirring, was always sensitive to the more speculative and philosophical aspect of theoretical physics, and he had been introduced to the foundational problems of quantum mechanics already in his years as a student, thanks to the presence of Schr\"odinger \cite{pieinter}. But it was only in 1964 that Pietschmann --together with his friend, the philosopher Gerhard Schwarz, who was working under Erich Heintel (1912-2000)-- established a novel dialogue between physics and philosophy, breathing new life into this Viennese tradition. Indeed, the two founded a discussion group called \emph{Philosophisch-Naturwissenschaftlicher Arbeitskreis} (PNA),\footnote{Apart from Pietschmann and Schwarz the following people were organisers and regular members of the PNA: Fritz Grimmlinger, Uwe Arnold, Ulrich Sexl, Karlheinz Schwarz, Dieter Klein and Günter Vinek.} a discussion group that focused on topics at the intersection of physics an philosophy, debated in front of an audience of students.\footnote{The history and the activities of the PNA have been recently collected in the book \emph{Philosiphysik} \cite{philosophysik}, from which we borrowed the title of the present paper.}  
Since the beginning, Pietschmann and Schwarz --at that time both young postdoctoral researchers in Vienna-- wanted to establish their PNA as a official university course. 
They individually approached their respective supervisors for their approval as professors. Walter Thirring himself had little interest in philosophy, but Pietschmann recalls that “Thirring always was  very open minded and said: 'yes, it is fine if I don't have to be involved'." \cite{philosophysik}. From the side of the philosophers, the response was somewhat more pessimistic about the success of the course. Schwarz recalls his mentor saying:
\begin{displayquote}
This is not gonna work Mr. Schwarz, because physicists don't understand the first thing about philosophy and it is not possible to teach them. But if you want to do that!  \cite{philosophysik}
\end{displayquote}
Despite this reluctance, the PNA took off as an official university course where students could get a grade by attending the panel discussions and handing in a paper. 


Soon after its foundation, another prominent participant joined the PNA, Pietschmann's colleague Roman U. Sexl. The latter was the nephew of the aforementioned Viennese Physicist Theodor Sexl, from whom Roman possibly learned about the fundamental problems of quantum physics \cite{pieinter}. Sexl completed his PhD, in 1961, under Walther Thirring and, after a period in the US at several different institutions, he came back to Vienna in 1967, where he became full professor of General Relativity Theory and Cosmology in 1971. His work on the production of particles by gravitational fields \cite{sexl68}, carried out with his Viennese colleague Helmuth Urbantke, pioneered the field of particle production in curved space time that would lead to Hawking's radiation a few years later, and  won him international fame. Moreover, Sexl became renowned for his involvement in physics education, and wrote several books both scholarly and popular science ones.
Sexl had a very dynamical intellectual style, as recalled by Bertlmann:
\begin{displayquote}
he was very broad in this. He was typically Viennese in this. Not just doing calculations, but also philosophy, culture, art. \cite{bertlmanninterview}
\end{displayquote} 
Indeed, since 1979, Sexl was the editor of the book series \emph{Facetten der Physik} (“Facets of physics") that promoted a genuinely interdisciplinary and open-minded approach to physics, as stated by the introductory editorial note that opens every volume of the series: 
\begin{displayquote}
Physics has many facets: historical, technical, social, cultural, philosophical and amusing ones. They can serve as essential and decisive motives for the pursuit of the natural sciences.
\end{displayquote} 
It is in that book series that Selleri, the main character of the Italian revival of FQM in the 1970s (see section \ref{hiddenvar}), published his “Die Debatte um die Quantentheorie" (\textit{The Debate over the Quantum Theory}) \cite{selleri1983} in 1983.  Sexl himself --together with Kurt Baumann-- published, in 1984, “Die Deutungen der Quantentheorie" (\textit{The Interpretations of Quantum Theory}) \cite{sexlbaumann}, which is a collection of original papers that aims to show the different interpretations of quantum formalism (they include papers by the Copenhagen school, as well as the alternative views of Schr\"odinger, Fock, Bohm, Bell and deWitt). Remarkably, Sexl's book shows an exceptionally modern understanding of the relevance of the interpretational problemes:
\begin{displayquote}
Because only mathematical formalisms that have an interpretation can be understood as a physical theory. [...] Even though, at some point in time, different interpretations lead to the same physical results, they let us anticipate very different developments and research directions  \cite{sexlbaumann}.
\end{displayquote}
Moreover, together with Peter Aichelburg, Sexl edited a collective volume in honor of Einstein in the occasion of the centenary from his birth. The book, entitled \textit{Albert Einstein: His influence on physics, philosophy and politics} \cite{einsteinbook}, collected contributions from distinguished physicists and philosophers.\footnote{Karl Popper was also invited to contribute but he did not manage to do it (letter from Popper to Roman Sexl on March 20th, 1978. Courtesy of Peter Achelburg).} Interestingly, it also contained a chapter authored by Nathan Rosen on the EPR paradox, wherein Einstein's collaborator stated his lack of confidence that quantum theory could be completed in terms of hidden variables, because this would mean to give up locality in the light of Bell's theorem.      

In the spirit of ``Facets of physics", the PNA covered a diverse palette of topics in their discussions, each of which usually took up several semesters. In the course of over 50 years, the PNA covered 12 broad topics in their discussions \cite{philosophysik}. It started with the subject of \emph{Space and time} (1964-1969), which was to become a common theme over the course of the following decades. Its continuation led to questions regarding \emph{Inertial and gravitational mass} (1969-1971, 2004-2006), where the assumptions leading to ideas such as cosmic inflation and the theory of the big bang were received critically by the philosophers of the PNA, who considered them rather arbitrary. This lead Pietschmann to introduce the distinction between predictive and descriptive theories, which he published in \cite{pietschmann2}. As a consequence, the discussion shifted toward methodology (\textit{Methodenproblem}) (1971-1978, 2004-2006). In the following years the PNA discussed an array of different topics, including \emph{the sociopolitical meaning of science and technology} (1979-1982, 1995, 1996),  \textit{Artificial intelligence} (1982-1985),  \emph{Quality and quantity} (1986-1992), \emph{Virtual reality} (1993-1995, 1998-1999), and \emph{fundamental aporias of Aristotle}.

In their meetings, they sometimes invited guests to contribute to the discussion, such as I. Birkhan, C.F. Weizsäcker, K.  Kumpf or R. Fischer. Of particular interest is the involvement of Karl Popper, who became a major influence on Pietschmann's thoughts. Sexl and Pietschmann, indeed, invited Popper to participate in the discussion of the PNA in October 1972, and kept a correspondence with him until 1980.\footnote{Karl Popper Archives, Box/Folder: 349/4, AAU, Klagenfurt (Austria)/Hoover Institution, Stanford (California).} Pietschmann, even published the paper \emph{The Rules of Scientific
Discovery Demonstrated from Examples of the Physics of Elementary Particles} \cite{pietschmann1}, wherein he used examples from theoretical particle physics --his main expertise-- to show that the methodological rules of Popper's falsificationism are ``actual tools rather than abstract norms in the development of physics", and that working physicists in fact use them in their scientific practice. Despite his influence on the PNA and the repeated invitations, there is no evidence that  Popper eventually ever attended their meetings.

The PNA certainly had a strong impact on the general landscape of physics in post-war Vienna, where the new developments of high energy physics, under the lead of Walter Thirring, had overcome the philosophy-oriented mood that had characterized the research in physics before WWII. In the particular case of Pietschmann, for example, the interaction with philosophers was to have lasting and deep consequences, and paved the way for his subsequent interest in foundational questions. This is witnessed by Aichelburg, who personally did not find the PNA particularly fruitful, but recalls that ``especially Pietschmann --Roman Sexl was more the practical type-- [...] somehow absorbed this philosophical vocabulary and somehow turned to the philosophical aspects.” \cite{aichelburg}. Also Zeilinger participated in at least one of the meetings of the PNA, but did not find it particular influential in his intellectual development. However, he maintains that the very existence of such a successful initiative was symptomatic of a remarkable openness towards foundational questions in the Viennese environment, which likely helped pave the way for the establishment of FQM in the following years.\footnote{Zeilinger also recalls that he regularly participated in an other monthly group of discussion between physicists and philosophers in those years, led by philosopher Michael Benedikt (1928-2012). We are thankful to Anton Zeilinger for sharing this in a personal communication on September, 17th 2021.}  The PNA was so successful as to last for over half a century  (a final meeting of the PNA took place in 2016).

\subsubsection{Physics education at the University of Vienna in the 1960s-1970s}

Before entering the core of our reconstruction about the first steps towards modern quantum foundations in Vienna, we deem it relevant to give an overview of the system of physics education at the University of Vienna in the 1960s and 1970s. In fact, as we will see in the next sections, it was in that period that some of the protagonists of the rebirth of FQM --such as Zeilinger and Bertlmann-- were students in Vienna.

As already recalled in the introduction, a peculiar aspect of the University of Vienna is that up until 1975 the Physics Institutes were part of the Faculty of Philosophy. This was not just an administrative matter, but had a real impact on the education of young physicists. Pietschmann recalls that at that time, in fact, physics students 
\begin{displayquote}
had to take five \textit{Rigorosen}, [final exams for achieving a doctorate] three in the main topic --in [this] case physics-- and two in philosophy. So [they] really had to study it and learn it, although the philosophers knew that they had to be very soft with scientists \cite{pieinter}.
\end{displayquote}
Zeilinger also recalls that the course he followed in philosophy to prepare for the \textit{Rigorosen} surely had an impact on his intellectual formation.\footnote{In particular, his two examiners in philosophy were Johann Mader (1926-2009) --whose course on the historical development to the fundamental concepts of philosophy (\textit{philosophische Grundbegriffe})
 he found very interesting-- and the logician Curt C. Christian (1920-2010). (Personal communication from Anton Zeilinger to the authors on September, 17th 2021).}
In general, this formal training in philosophy is likely to have played a considerable role in forming the interest towards foundational problems at the University of Vienna.

Moreover, the curriculum was exceptionally flexible, granting students a customized training of their choice.\footnote{Anton Zeilinger recalls that only one advanced laboratory course (\textit{Praktikum}) and two theoretical seminars were mandatory. (Personal communication from Anton Zeilinger to the authors on September, 17th 2021).} This allowed students who approached physics with a genuine curiosity for fundamental questions --which presumably is a rather common motivation to study physics-- to foster this curiosity, in contrast to many strict formal curricula. On this note, Bertlmann remembers: “When I think back it was like in paradise, I must say. You only had to chose your thesis advisor and he told you what you had to study." \cite{bertlmanninterview}. Indeed, until the reform of 1975, there was a single cycle degree program (which if completed would grant a PhD) without mandatory courses or exams to take, except the final Rigorosen. Also Zeilinger, who studied physics in Vienna in exactly the same years of Bertlmann, recalls:
\begin{displayquote}
when I started to study physics and mathematics at the University of Vienna in 1963, there was no fixed curriculum at all. One was essentially free to choose the topics according to one’s liking. Only at the end, one had to pass a rigorous examination and present a PhD thesis. This resulted in me taking not even a single hour of quantum mechanics, but I learned it all from textbooks for the final exam \cite{zeilinger2017}.\footnote{Pietschmann, who introduced the habit for experimentalists to take one of the \textit{Rigorosen} in theoretical physics and vice-versa, was one of  Zeilinger's examiner on the theory part \cite{pieinter}. Zeilinger himself had asked him to be examined on quantum physics, which he studied on a book at that time recently published by a Professor of Theoretical Physics at the TU, Otto Hittmair \cite{hittmair}. (Personal communication from Anton Zeilinger to the authors on September, 17th 2021).}
\end{displayquote}

For what concerns the teaching of quantum mechanics, we have already stated that this was taught for the first time by Walter Thirring, from 1959, in his course of theoretical physics. In Thirring's course --as we mentioned already, he was more fascinated by the mathematics than by the philosophical questions behind physical theories-- students were not introduced to any foundational problems, such as the EPR paradox or the measurement problem in quantum mechanics.  During the  years 1968-1971, when Thirring moved to CERN, it was Pietschmann who replaced him in teaching theoretical physics, which at that time was a two-year course divided in four parts: mechanics, electrodynamics, quantum mechanics and statistical physics \cite{pieinter}. However, despite his full involvement in philosophical discussions, leading the PNA (see section \ref{PNA}), and his knowledge of the conceptual issues of quantum theory,  also in his course topics pertaining to FQM where not introduced. Nevertheless, he reorganised the structure of the course in a unconventional but thoughtful way. He would directly start by introducing thermodynamics, immediately turning to quantum mechanics afterwards, because “if you start from classical mechanics, they [students] will never grasp the idea of quantum mechanics", Pietschmann recalls today \cite{pieinter}. The same lack of foundational topics applies as well to the courses held by Roman Sexl, who also used to teach quantum mechanics in the following years. Yet, despite \textit{Facetten der Physik} was publishing whole volumes on FQM, and their involvement in the PNA, neither Pietschmann nor Sexl seemed to have introduced their interest towards foundational problems in their teaching throughout the 1970s and even the 1980s. Bertlmann, who attended their courses himself, recalls:

\begin{quote}
    The lectures of Pietschmann and Sexl were very good, very modern, but no density matrices and of course, no Bell's Theorem. Although Sexl was writing about Bell's Theorem in a book at that time, so he knew about it. But he never mentioned it in the course. It was an official line not to do it \cite{bertlmanninterview}.\footnote{Also Dieter Flamm (1936-2002) --the son of the physics professor Ludwig Flamm and of the youngest daughter of Ludwig Boltzmann, Elsa-- lectured on theoretical physics but his lectures were extremely formal and did not touch upon foundational aspects at all \cite{bertlmanninterview}. However, around 1993, he held a seminar course on the foundations of quantum mechanics, attended only by a few students.  (We are thankful to \v Caslav Brukner for pointing this out in a personal communication on September, 16th 2021).}
\end{quote}
One can thus see a dichotomy between the interest towards foundational and philosophical aspects of physics, that physicists such Pietschamann or Sexl were extensively cultivating, and their appearance as professor of the Faculty of Physics, especially in their teachings. In this respect, Aichelburg maintains: “if there was at that time some discussions on fundamental physics, I didn’t know this.” \cite{aichelburg}. We regard this attitude of keeping these interests somewhat private as a cultural product of the age of the “shut up and calculate", as Bertlmann puts it, “it was an official line".


\begin{figure}[ht!]
  \centering
  \includegraphics[width=\linewidth]{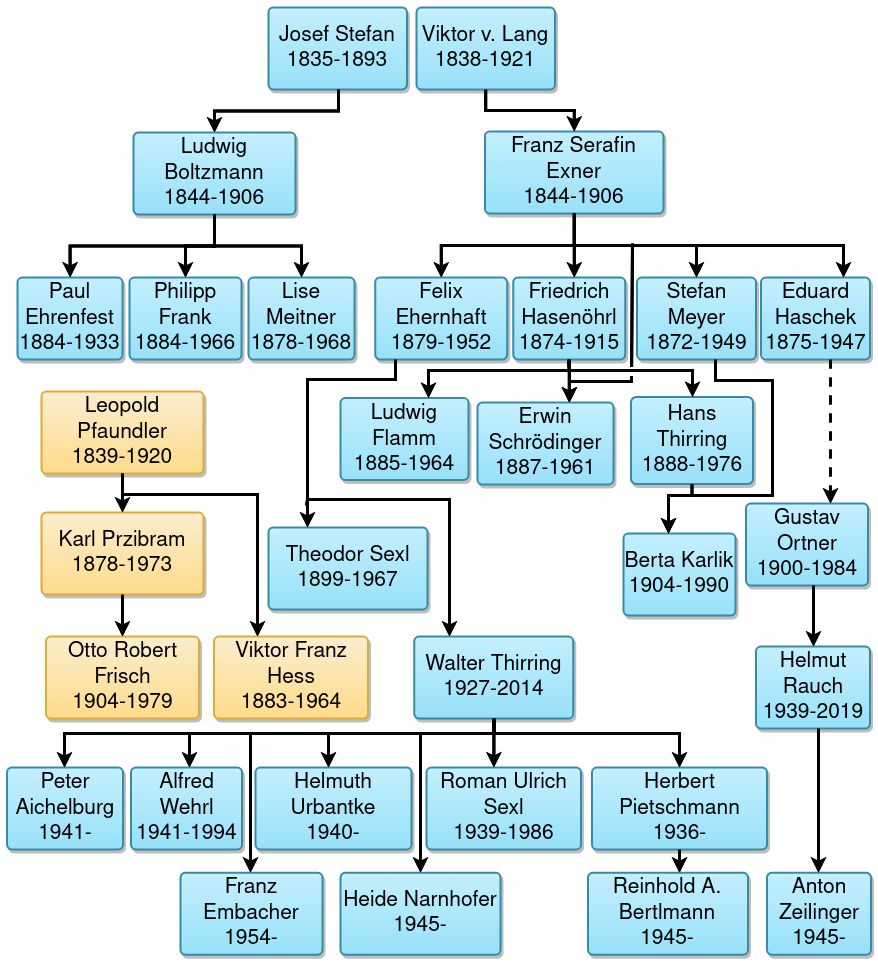}
  \vspace*{-1.5mm}
  \caption{{\footnotesize Academic “family tree" of the relevant personalities mentioned in this work. The links between them indicate a supervisor-doctoral student relation (or in the rare cases in which it was not possible to certify this, an acknowledged major influence). The link between Eduard Haschek and Gustav Ortner is dotted because we have inconclusive evidence as to weather there was a strong connection between them. The yellow cluster indicates an independent branch originating from Leopold Pfaundler (who actually never worked in Vienna).}}
  \label{fig:familytree}
\end{figure}
\section{The rise of quantum foundations in Vienna}

\subsection{Experimental FQM with neutrons: Rauch and Zeilinger}
\label{rauch zeilinger}

The \emph{Atominstitut} is located just a few kilometers south-east of the physics building of the University of Vienna, and it has been another major venue for the rebirth of the Viennese foundational research. This institute, initially called \emph{Atominstitut der {\"o}sterreichischen Universit\"aten}, was established in 1958 as an inter-university institution --between the \emph{Technische Hochschule} (which was renamed  Technical University of Vienna in 1975, hereinafter TU) and the University of Vienna-- to host the TRIGA Mark II research reactor.
Helmut Rauch, one of the protagonists of the rise of FQM in Vienna was based there. He had started his studies at the TU in 1957. About his study years, he recalled that “quantum mechanics was taught in the standard form, that means Copenhagen as it is, and there were no broad discussions about that." \cite{rauch interview}. It is worth noting that the TU never had a Faculty of Philosophy, which might be on of the reasons why the connections to philosophy were even weaker then at the University of Vienna. However, also there one can find some exceptions:  Walter Glaser (1906-1960), professor of Theoretical Physics at the TU  since 1956, was said to have strong interest towards the foundations of quantum mechanics, and to oppose the Copenhagen interpretation. Furthermore, Rauch recalled that occasional visitors would come to the TU to give talks about Bohmian mechanics, and that Karl Popper visited the institute several times around the mid-1960s, because he was very interested in matter-wave quantum phenomena \cite{rauch interview} (see Section \ref{hiddenvar}).

Rauch completed his doctoral research in 1965 at the Atominstitut with a thesis entitled “Anisotropic $\beta$-decay after the absorption of polarized neutrons", under the supervision of Gustav Ortner (1900-1984).\footnote{Ortner had become very influential during the Nazi period, when he served as director of the Radium institute of the Academy of Science \cite{kerne}. He was suspended from office at the end of the WWII, when he emigrated and became professor of physics at the University of Cairo (Egypt) until 1955. After his return to Austria, in 1960 he became Professor of Nuclear Physics at the TU.} Ortner was one of the founders and the director of the Atominstitut, and according to Rauch he “was very open and [...] interested of doing fundamental physics which could be published in reasonable journals." \cite{rauch interview}. In March 1972, Rauch was appointed full professor of Experimental and Neutron Physics at the TU, and in the following years he was to be recognised as one of the most renowned experts on neutron interferometry worldwide.
Indeed, following the success of the X-ray interferometry on silicon crystals achieved by Ulrich Bonse (1928-) and Michael Hart at Ithaca College (NY) \cite{bonse1965}, Rauch's group obtained a pivotal result in 1974, becoming the first one to realize a silicon crystal appropriately manufactured to implement a (Laue-type) interferometer between coherent beams of neutrons \cite{rauch1974}. This gathered some international attention, including the expert on nuclear reactors for research on neutrons, Heinz Maier-Leibnitz (1911-2000), who invited Rauch and his group to install the silicon crystal at the Institute Laue–Langevin in Grenoble (France), where many of the following experiments took place. 

For what concerns the FQM, it is worth mentioning that Rauch started a seminar course around 1968-69 on neutron interferometry, wherein conceptual and foundational aspects were discussed at length.\footnote{We are thankful to Anton Zeilinger for sharing this is a personal communication on September, 17th 2021.}

In 1968, also Anton Zeilinger --at that time a student at the University of Vienna-- joined Rauch's group, wherein, in 1971, he completed his doctoral dissertation, entitled \emph{Neutronendepolarisation in Dysprosium-Einkristallen} (“Neutron Depolarization in Dysprosium Single Crystals"). 

Zeilinger recalls that his interest on foundations of physics was easily accepted and grew within the group of Rauch at the Atominstitut, feeling that he had escaped the reluctance with which foundational issues were still regarded at the Faculty of Physics of the University of Vienna.\footnote{Personal communication from Anton Zeilinger to the authors on September, 17th 2021.} 
Zeilinger was particularly involved --together with Gerald Badurek, who was to become one of the successors of Rauch in the field of neutron interferometry-- in the first experiment of foundational nature conducted with the new neutron interferometer: the verification of the sign flip of the wave function of a spin-$1/2$ particle. In fact, quantum theory predicts the peculiar effect that the wave function of a spin-$1/2$ particle, such as a neutron, changes its sign when undergoing a rotation of $2\pi$ \cite{rauch1975}. While in classical physics no change can be expected under rotations of $2\pi$, even in quantum mechanics this effect had been long considered undetectable since observables in quantum theory are quadratic in the wave function. However, by means of the new technique of neutron interferometry, Rauch and his collaborators where the first ones to experimentally demonstrate this genuine quantum effect.\footnote{For a review on Rauch's foundational experiments on neutron interferometry see \cite{rauch2012}.} 
That experiment proved to be the real turning point in Zeilinger's career, not only for the scientific relevance of the result, but also because this offered the first opportunities for his involvement in the (at that time still rather small and unconventional) community of physicists concerned with quantum foundations. In fact, on April 18-23, 1976, an international workshop on FQM, “Thinkshop on Physics”, took place in Erice (Italy), organised by the “Ettore Majorana Foundation and Centre for Scientific Culture". Directed by John Bell and Bernard D'Espagnat (1921-2015), the event gathered 36 scholars from 25 institutions located in 9 different countries (see \cite{freire2014}).
This workshop had a pivotal international resonance for the field of FQM, it “was seen by some physicists as the turning point in the acceptance that quantum nonlocality was indeed a new physical effect.” \cite{freire2006}. John Clauser (1942-), a pioneer of research on Bell's inequalities, pointed out that “the sociology of the conference was as interesting as was its physics. The quantum subculture finally had come out of the closet.” \cite{clauser1992}.  
There, Zeilinger had his first encounter with quantum entanglement and Bell's inequalities, topics that  were to become his main research interests and he was to give decisive contributions to their theoretical understanding and especially to their experimental realizations and applications. Zeilinger himself recalled that occasion, many years later:
\begin{quote}
    The topic of this workshop was exactly experiments on the foundations of Quantum Physics. I went there and reported on our newest experimental results in neutron interferometry. I should mention that this was my first real encounter with the international scientific community. There, I heard for the first time about Bell's theorem, about the Einstein-Podolsky-Rosen paradox, about entanglement, and the like. [...] This meeting turned out to be very crucial in my life.  There, I met a number of colleagues for the first time, some of whom later became personal friends. These include Mike Horne, Abner Shimony and also Val Telegdi. Mike Horne and Val Telegdi then helped me to get to MIT and work with Cliff Shull at the neutron diffraction laboratory there. \cite{unspeakable}.
    \end{quote}

Indeed, in Erice Zeilinger became friend with the renown experimental particle physicist Valentine Telegdi (1922-2006) who then wrote to Viktor Weisskopf at the  Massachusetts Institute of Technology (MIT), and thanks to whom, a few weeks later, Zeilinger was invited by Clifford Shull (1915-2001) --a pioneer of neutron diffraction, for which he was to win the Nobel Prize in 1994--  to join  MIT. Zeilinger spent the periods from summer 1977 to end of 1978, and from 1981, when he became associate professor there, to 1983 at MIT. During that time, he kept working on neutron interferometry, discussing related foundational problems with the theoretician Michael Horne (1983-2019).\footnote{As early as 1969, Horne had been coauthor of a pivotal paper on an experimentally testable version of Bell's inequalities, know as Clauser-Horne-Shimony-Holt (CHSH) inequality \cite{chsh}.}  

It should be stressed, however, that Zeilinger's research did not deal with entanglement and non-locality (nor did he discuss these topics with Horne) until 1985, when they found out that a conference on the occasion of the fiftieth anniversary of the Einstein-Podolski-Rosen (EPR) paper was to be held in July that year, in Joensuu (Finland). It was precisely for having something to present at that conference that Zeilinger and Horne formulated the first experimental proposal for violating Bell's inequalities using entanglement between external degrees of freedom (as opposed to internal ones, such as spin) \cite{gilder}. That work  \cite{finland}, published in the proceedings of the conference, was the first paper of Zeilinger on matters of entanglement and non-locality. The collaboration with Horne was soon after joined by theoretician Daniel Greenberger (1932-), whom Zeilinger had first met at a conference in June 1978 at the Institute Laue-Langevin in  Grenoble, and who was a regular visitor of Shull's group at MIT. In 1987, when Zeilinger was back in Vienna as associate professor, Greenberger joined him at the Atominstitut, as a Fulbright fellow. During that period, together with Horne (collaborating at-a-distance from Boston), they were the first to tackle the problem of tripartite entanglement, which allowed them to formulate a prime theoretical result in FQM: a stronger form of Bell's theorem -- the  GHZ theorem-- where the quantum correlations are no longer probabilitstic but deterministic, so-called perfect correlations.
 \cite{ghz}. 

Zeilinger was prominently an experimentalist, so his aim became to realize some of these theoretical proposals involving multipartite entanglement in a laboratory. However, he soon realized that neutrons are not suitable to create entangled states, so he switched to photonic entanglement, by applying the newly developed techniques for creating entangled pairs of photons through the effect called spontaneous parametric down-conversion; (timewise, this coincided with Zeilinger's moving to the University of Innsbruck as full professor of Experimental Physics, a position which he held between 1990 and 1999).

In the following years, Zeilinger became increasingly recognized as a world's leading figure in FQM, quantum optics and quantum communication. Reconstructing the scientific achievements of Zeilinger in these fields goes beyond the scope of the present paper; detailed accounts can be found in \cite{gilder} and \cite{freire2014}, whereas for a personal recollection of Zeilinger's experiment in quantum optics we refer to \cite{zeilinger2017}. Here, we limit ourselves to retrace a few selected contributions in his career: In 1993, together with the Polish physicist Marek \.Zukowski (1952-), Horne and Artur Ekert (1961-), Zeilinger proposed a protocol to achieve \textit{entanglement swapping}, which allows to entangle two particles that have never interacted \cite{Z0}; in later years he also demonstrated this phenomenon experimentally \cite{Z0.5}. In 1994, Zeilinger and collaborators showed that any discrete finite-dimensional unitary operator can be constructed in a laboratory using only linear optical elements \cite{recketal}. This is a major breakthrough because it formally justifies the use of table-top optical experiments to implement basically any quantum effect. In 1996, Zeilinger's group achieved the first experimental realization of the \textit{dense-coding}, which demonstrates an increase in the channel capacity if quantum resources are used in communication \cite{Z1}. The following year, they realized the first experimental \textit{quantum teleportation} \cite{Z2}, which allows to “teleport"  the quantum state of an unknown system (i.e. to send the state without having it travel between the sender and the receiver); this led in more recent years to experiments of increasing ambition, such as teleporting a quantum state over 143 km \cite{Z3}. Moreover, Zeilinger experimentally realized the GHZ tripartite entangled states \cite{Z4}; achieved the first violation of Bell's inequalities closing the communication loophole \cite{Z5} and, recently, one of the first loophole-free Bell's tests \cite{Z6}.

Moreover, on Zeilinger's (amongst others) initiative, in 2003 the “Institute for Quantum Optics and Quantum Information" (IQOQI) of the Austrian Academy of Sciences was founded, with two independent sites in Innsbruck and Vienna, respectively, to promote the research in the developing fields of theoretical and experimental quantum optics and quantum information.\footnote{See \href{https://www.iqoqi-vienna.at/the-institute/history}{https://www.iqoqi-vienna.at/the-institute/history}. Last accessed on 30.06.2021.}    

\subsection{The influence of the “quantum dissidents" on Vienna's FQM}\label{sec:exernal}

As already recalled, in the post-war period, the research on FQM had encountered a tremendous setback, with the only exception of a few physicists that the historian of physics Olival Freire Jr. has named the “quantum dissidents", because they challenged the standard view according to which “foundational issues were already solved by the founding fathers of quantum physics" \cite{freire2014}.
It ought to be remarked that the majority of these dissidents --such as David Bohm (1917-1992), John Bell, Hugh Everett (1930-1982), Jean-Pierre Vigier (1920-2004) and Franco Selleri-- who effectively helped a great deal bring FQM out of oblivion, were motivated by a staunch commitment to realism, against the widespread Copenhagen interpretation of quantum mechanics. Curiously enough, since quantum foundations mainly merged into the field of quantum information science (which became increasingly popular in the 1990s) the new trend of studies on FQM has greatly distanced itself from the realist velleities that originally motivated the quantum dissidents. In particular, the prominent school of thought initiated by Zeilinger in Vienna, which indeed came to foundations rather late and helped develop quantum information, mainly supports a non-realist “Neo-Copenhagen" interpretation, where the epistemic concept of information plays the main role \cite{zeilinger1999}. Nevertheless, in this section we argue that some of the aforementioned quantum dissidents helped popularize FQM in Vienna, and have likely played a relevant role in smoothing the path towards the rise of the successive “Zeilinger era" of FQM.  

\subsubsection{The hidden variable program: Selleri, Vigier and Popper}
\label{hiddenvar}
The initiator and main actor of the Italian rebirth of the interest towards FQM in Italy was the theoretician Franco Selleri.\footnote{For a detailed reconstruction of the history of FQM in Italy and in particular on the role played by Selleri see \cite{baracca2017, delsanto2020, romano2020}.} He was an unconventional physicist who abandoned a secure career in particle physics --where he had already given remarkable contributions-- to start a decades-long battle against the Copenhagen interpretation of quantum theory. As early as 1969, he published a first paper \cite{selleri1969} --praised by the Nobel laureate Louis de Broglie (1892-1987)-- where he adopted a “hidden variable" approach which associates an objectively existing wave to each quantum particle (although there could exist “empty" waves even without particles). Worth mentioning is that Selleri's scientific positions were explicitly rooted in his radical left-wing, materialistic credo,\footnote{Indeed, there was a tradition of left-wing physicists who, supporting materialism,  opposed the Copenhagen interpretation because considered anti-realist. This was the case of the early motivation for Bohm's alternative interpretation \cite{freire2019} and for the work on hidden variables of the French physicist Jean-Pierre Vigier \cite{besson2018}.} and that in the 1970s a remarkably large group of young physicists --which partly included Giancarlo Ghirardi (1935-2018), who was to give important contributions to FQM-- joined his research program. 

Selleri and his school took part in the Erice workshop of 1976, and more than likely it is there that he started to be interested in experimental tests of some of his ideas on FQM using neutrons. Many years later, he would recall:
 \begin{quote}
I have done a lot to promote neutron interferometry as a tool to study fundamental questions. [...] I was particularly connected with Helmut Rauch at Vienna and Rauch is a practical man, less philosophically inclined than Zeilinger but a very concrete physicist. And I believe that neutron interferometry shows very clearly that you have wave particle duality again.  \cite{freire2003}
 \end{quote}

Indeed, after 1982 when the first experiments on Bell's inequality were conducted with photons by Alain Aspect  (1947-) \cite{aspect} --which confirmed the violation predicted by quantum theory and thus jeopardised most of the hidden variable programs (because the alleged hidden variables would need to be explicitly non-local)-- neutron interferometry became for many the hope to save hidden variables.\footnote{Stuart Freedman (1944-2012) and John Clauser had already experimentally violated Bell's inequalities in the early 1970s, but  Aspect's experiment is considered to be the first one to have properly addressed the problem of locality (i.e. without the so-called locality loophole). In fact, the periodicity of the measurement settings in Aspect's experiment, cast doubt on the fundamental validity of his experiment too, and the first reliable experiment closing the locality loophole is considered by some to be Zeilinger's \cite{Z5}.} In March 1983, at the conference organised by the \emph{Association Vaudoise des Chercheurs en Physique} in Montana (Switzerland), Bell himself --who formulated the inequalities named after him to empirically discriminate between \emph{local-realism} and quantum mechanics, inasmuch as he was a supporter of the former--  had a very long discussion with Rauch, explaining him that “it would be important to make it [Bell's experiment] with matter waves, because electromagnetic waves are in principle all over [...]."  \cite{rauch interview}.

This view was shared not only by Selleri, who hoped to be able to use experiments of neutron interferometry also to demonstrate the existence of his postulated empty waves, but also by his colleague, the French theoretical physicist Jean-Pierre Vigier. He was a former student of de Broglie and the collaborator of Bohm, who probably more than anyone else had popularized hidden variable theories in Europe \cite{besson2018}. Vigier visited Rauch's group several times, both at the Atominstitut in Vienna and at the Institut Laue–Langevin in Grenoble, where he tried to persuade Rauch to conduct experiments that could allegedly disprove the Copenhagen interpretation. Indeed, Rauch recently stated: 
\begin{quote}
many of our experiments were afterwards really stimulated by some strange ideas about how to distinguish between different interpretations. [...] Also Vigier was a fighter for that. But [...] we
were fortunately always very careful in our discussion. We never mentioned that it disproves quantum mechanics [...] \cite{rauch interview}.
\end{quote}

In the early 1980s, Vigier has also been the main influence on the over-80-years-old philosopher Karl Popper --who had been engaging with problems of FQM since 1934 and as such should be considered a fully-fledged “quantum dissident" \cite{delsanto2019}-- to actively engage again with the community of quantum physicists. Indeed, at a conference in Bari, organised by Selleri himself, Popper presented a “simplified version of the EPR experiment", which by means of measurements on an entangled pair of particles supposedly would have discriminated between a realist interpretation and the Copenhagen one. Selleri had arranged that some experimentalists, including Rauch, would be present with the hope to persuade someone to perform Popper's proposed experiment \cite{delsanto2018}. Consequently, Popper visited Rauch in Vienna in March 1984. Rauch wrote to Popper that he had found his proposal very interesting and invited him to give a talk at the \emph{Chemisch-Physikalische Gesellschaft } (“Chemical-Physical Society") in Vienna, of which Rauch was the chairman.\footnote{Letter from Rauch to Popper on May 2, 1984. Karl Popper Archives, Box/Folder: 341/9, AAU, Klagenfurt (Austria)/Hoover Institution, Stanford (California).}
In the end, Popper's experiment was carried out only in 1999, after his death, but in hindsight it turned out that Popper's proposal was not capable of discriminating  between different interpretations of quantum physics (see  \cite{delsanto2018}). On the other hand, Bell's experiments with neutrons did not work because of the technical difficulties to realise a source of entangled pair of neutrons \cite{rauch interview}. However, the interaction between these unorthodox thinkers and the group of Rauch stimulated the foundational character of the  experiments with neutrons, as explicitly acknowledged by Rauch, and more generally may have contributed to open to discussions on FQM in Vienna.

Besides neutron interferometry, Selleri’s influence in Vienna took also a different path, one which could have potentially left deeper marks since this time it involved students too. Selleri had been good friend with Pietschmann for many years (they had met at CERN in Geneva in 1960-61, when Selleri was still working on particle physics) and, at least from the beginning of the 1980s, started a fruitful intellectual relationship with Sexl too. In fact, it most likely was the interest of Sexl and Pietschmann towards foundational questions that caused Selleri to spend a leave of absence at the institute of Physics in Vienna. His visit lasted from winter 1980 to spring 1981 and was followed by further regular visits until at least 1985.\footnote{In a letter dated August 5, 1890, Pietschmann informs Selleri of having succeeded in getting financial support, starting from October 15 of the same year. We were able to find evidence of five more official visits between October 1981 and January 1985 (Courtesy of Luigi Romano,  University of Bari).} In Vienna, indeed, Selleri used to discuss problems of FQM mainly with Pietschmann, Sexl, and Rauch; Also Bertlmann, still a young researcher at that time, remembers about Selleri's visits, but also gives us a taste of how unpopular the topic of FQM was: he “did not have contact with him [Selleri] on this subject. This was not allowed, so to say" \cite{bertlmanninterview}. Therefore, it is rather surprising that, during his period in Vienna, Selleri was actually allowed to give lectures on FQM.  Selleri himself remembered that Sexl “followed [his] lectures in Vienna and invited [him] to write a book” on their topic \cite{freire2003}, which appeared in 1983 as a volume \cite{selleri1983} in the already mentioned series \emph{Facetten der Physik}, edited by Sexl himself (see section \ref{PNA}).\footnote{The book was reprinted translated in various languages in the following years, including a Spanish version which included a preface by Popper \cite{delsanto2018}.} The contents of this book reflect the unconventional palette of topics that Selleri had covered in his lectures in 1980.  It encompasses whole chapters devoted to problems such as the completeness of quantum theory and hidden variables, wave-particle duality from a realist point of view (which includes also experiments on neutron interferometry), the Einstein-Podolsky-Rosen Gedankenexperiment and Bell's inequality, as well as a final chapter on “Experimental Philosophy".  

 Moreover, while in Vienna, Selleri wrote a paper \cite{selleri1980} in which he proposed a model that allegedly disproved the fact that the factorability of probabilities for outcomes of experiments conducted at distant locations is sufficient to derive Bell's inequalities, as proposed by Clauser and Horne \cite{clauserhorne}.\footnote{The paper was written with the help of one of his pupils, the philosopher of physics Gino Tarozzi. The authors are thankful to Gino Tarozzi for sharing his testimony on his visit to Vienna in a personal communication to one of the authors (F.D.S.) on February 12, 2018.} This proposal found the praise of Popper --who coined the term “universality claim" to refer to the result of Clauser and Horne (see \cite{delsanto2018})-- and had a certain following throughout the 1980s. However, it turned out to be untenable because it was based on a misconception (see, e.g. \cite{cushing} for a short overview). 
 
 Selleri is also known for having been among the first proposers of protocols to achieve superluminal communication, based on (a misconception  of) quantum entanglement (see \cite{delsanto2020}). He had proposed this independently of the works of Nick Herbert (1936-), who had promoted the same idea in the US (see \cite{kaiser}). It was in Vienna that Selleri found one such scheme for, allegedly, achieving faster-than-light communication. This, however, assumed the possibility for a laser to emit longitudinal modes all with the same, yet unknown, circular polarization. Pietschmann remembers that Selleri entered his office jubilantly and presented this result to him and Walter Thirring. The proposal had no formal mistake, but Pietschmann replied: “Absolutely great, you have just proven that it is impossible to make a laser which lases linearly polarised light in the same way", and Thirring agreed \cite{pieinter}. Indeed, the impossibility of preparing copies of unknown quantum states was to be proven in full generality soon afterwards (actually also as a reaction to Nick Herbert's proposal \cite{kaiser}), and became know as the “no-cloning theorem", on which the entire field of modern quantum cryptography relies.
 
 Interestingly, Selleri and Zeilinger wrote a paper together in 1988 \cite{zeilingerselleri}, despite their opposite views on FQM. Therein, they analysed a local deterministic hidden variable model that would determine whether a photon is detected or not in an EPR-like experiment. This model is empirically distinguishable from quantum mechanics, but this was at that time out of experimental reach, due to the low efficiency of photon detectors. The disagreement between the two authors is made manifest in the last sentence of the paper: “the difference in the expectations of the present authors whether this will happen or not is indicative of the diversity of opinion among physicists at large." \cite{zeilingerselleri}. On this note, Zeilinger today recalls: “I saw the paper as a mathematical exercise out of curiosity, for Selleri it was important for his realist view."\footnote{Email from Zeilinger to one of the authors (F.D.S.) on 17.09.2021.} As a matter of fact, all experiments so far performed keep confirming quantum mechanics.
 
 \subsubsection{John Bell's liegeman in Vienna: Reinhold Bertlmann}
 As already recalled, the quantum dissident \emph{par excellence}, namely the physicist that more than anyone else changed the understanding of modern FQM, was John Bell. He also played a role in Vienna, mostly (but not only) indirectly through the Austrian physicists Reinhold Bertlmann.\footnote{Bertlmann has published in recent years a series of papers \cite{bertlmann2002, bertlmann2014, bertlmann2017, bertlmann2020} with his personal recollection of John Bell and the latter's impact both on particle physics and on FQM. For an excellent scientific biography of Bell, see \cite{whitaker}.} The latter started his studies in physics at the TU in 1964, but moved to the University of Vienna two years later, where he completed his dissertation in particle physics in 1974 under the supervision of Pietschmann. During these years, Bertlmann was active in the political left-wing student movements, which were particularly sensitized against the involvement of physicists in military research (and more specifically in the Vietnam War). It was in that period  that Bertlmann's leftist attitude of protest led him to establish a curious wearing habit that, as we shall see, was to change his scientific life some years later. Bertlmann's himself recollects:
\begin{quote}
my way of looking was a protest against the bourgeoisie, but not only this. I don't know how it actually happened... One day I changed my socks. I thought: “why do all people choose the socks of the same colour for both feet, why is this?” So this for me was like a sheep-effect. And then I changed this. [...] I don't know the day exactly when I started, must have been in the '60s, but from that day on I never had socks of the same colour, no day. \cite{bertlmanninterview}
\end{quote} 

In 1977, Bertlmann had the opportunity to work for nine months in the Soviet Union, at the “Joint Institute for Nuclear Research" in Dubna. It should be recalled that Austria declared itself “perpetually neutral" in its new Constitution of 1955 (hence, it did not belong to either NATO, or the Warsaw Pact), so it enjoyed a particular status that allowed scientific collaborations to develop across the \textit{Iron Curtain}. There, Bertlmann was for the first time exposed to alternative interpretations of quantum mechanics, in particular to the realist-materialist position of  Dmitrii Blokhintsev (1908-1979) who had written a renowned book (within the Soviet Union) on FQM  \cite{dmitrii}.

In April 1978, Bertlmann moved to the Theory Division of CERN in Geneva, where he soon made the acquaintance of Bell \cite{bertlmann2017}. They started a close friendship and a prolific collaboration on topics of particle physics, notably on \emph{quarkonium}, i.e., bound states of quark-antiquark pairs (their first joint paper is \cite{bell bertlmann}, see Ref. \cite{bertlmann2017} for a bibliography).   
Bell was a recognized authority in the field of theoretical particle physics and accelerator physics --he was called “the Oracle of CERN" \cite{bertlmann2002}-- but he never discussed his paramount results on FQM (Bell's inequalities had been published as early as 1964! \cite{bell})  with his colleagues in CERN, nor did he discuss it with Bertlmann, who recalls: “John [...] never mentioned his quantum works to me in the first years of our collaboration. Why? This I understood later on, John was reluctant to push somebody into a field that was quite unpopular at that time." \cite{bertlmann2017}

The first direct contact between Bell and the University of Vienna occurred in May 1980, when he was invited by Walter Thirring as “Schr\"odinger Guest Professor” for about ten days. On that occasion, Bell gave three lectures: it must have been quite surprising for Thirring to figure out that only the first one was about particle physics, “On the role of duality for bound states in QCD” (on May 19), whereas the other two focused on his (at that time) virtually unknown work on FQM, “Assuming the Schr\"odinger equation is exact” (on may 21), and “On locality in quantum mechanics” (on May 22).\footnote{Peronal communication from Bertlmann to one of the authors (F.D.S.) on October 26, 2016.} During the last talk, Bell made a joke on the fact that EPR pairs behave in the same fashion of Bertlmann's socks. This triggered the laughs of the audience, but nobody, including Bertlmann, thought more about it \cite{whitaker}. However, on September 15, 1980, the joke came back into Bertlmann's life, launching him, willing or not, “out of the blue into the middle of the quantum debate"  \cite{bertlmann2014}. On that day, in fact, Bertlmann was at the  Institute of Physics in Vienna, which he had returned to in July, when his colleague Gerhard Ecker ran to him with a paper in his hands, shouting: “Reinhold look--now you’re famous!” \cite{bertlmann2017}. This was a pre-print authored by Bell, transmitted from CERN, whose title was “Bertlmann’s socks and the nature of reality” \cite{socks}. The paper was a restatement of Bell's inequalities, likely motivated by the fact that Aspect was at that time conducting his experiments. It explained in detail the fundamental difference between “spooky" quantum correlations and classical ones, exemplified by the (anti-)correlated colors of Bertlmann's socks. Indeed, the paper begins with:   
 \begin{quote}
The philosopher in the street, who has not suffered a course in quantum mechanics, is quite unimpressed by
Einstein-Podolsky-Rosen correlations. He can point to many examples of similar correlations in every day life.
The case of Bertlmann’s socks is often cited. Dr. Bertlmann likes to wear two socks of different colours. Which
colour he will have on a given foot on a given day is quite unpredictable. But when you see (Fig. 12) that the first
sock is pink you can be already sure that the second sock will not be pink.
[...] And is not the EPR business just the same ?”   
\cite{socks}
 \end{quote}

Bertlmann recalls that the paper pushed him abruptly into the field of FQM: “I had to study because everyone was addressing me as the most expert in the world, and I knew nothing!" \cite{bertlmanninterview}. And so he did, and finally engaged in many discussion on FQM  with Bell himself, also together with Jun John Sakurai (1933-1982), who was in CERN at the beginning of the 1980s --until his untimely death-- to write his famous book on quantum mechanics \cite{sakurai}, which was the first manual for students to include Bell's inequalities. Indeed, throughout the 1980s, Bertlmann became a main actor in popularizing Bell's results on FQM both in Vienna and internationally. He started being invited to give talks about Bell's inequalities and quantum non-locality, especially by physics groups that never dealt with FQM, such as experimental and particle physicists: In 1984, invited by Othmar Preining (1927-2007) at the Institute of Experimental Physics in Vienna, he gave the talk “\emph{Bell’sches Theorem}” (“Bell's theorem"); in 1987 he presented the lecture “Bell’s theorem and the nature of reality” firstly again at the University of Vienna, and in the same year --invited respectively by Heinrich Leutwyler (1938-) and Hans Günter Dosch (1936-)-- in Bern (Switzerland) and in Heidelberg (Germany).
Moreover, at the end of 1986, Bertlmann persuaded Walter Thirring to organize a  conference on quantum foundations at the University of Vienna, with an emphasis on Bell’s Theorem. The “Schrödinger Symposium”, took place the following year and, naturally, Bell was invited as a speaker. He gave a talk  on “Schrödinger’s cat” (on September 17, 1987), and also took part in the panel discussion (together with Zeilinger), announcing therein his notorious list of “words that should be forbidden in serious discussion”, such as “system", “apparatus", “observable", “measurement" (see \cite{bertlmann2020}).\footnote{Bell published this list in August 1990 in his paper “Against 'measurement'" to include “environment", “macroscopic", “microscopic", “reversible", “irreversible" and “information".\cite{Bell1990}} 

Throughout the 1980s --mainly thanks to the experimental violation of Bell's inequalities, but also thanks to the popularization of this topic by part of the quantum dissidents and other younger scholars-- Bell's theorem became increasingly recognized as “proper physics". 
On the occasion of Bell's sixtieth birthday, in 1988, Bertlmann wrote the paper “Bell's theorem and the nature of reality" \cite{bertlmann86}, whose pre-print he sent to Bell, who enjoyed it very much \cite{whitaker}, but also to Abner Shimony (1928-2015), another of the early experts in Bell's theorem. The latter, who got also James Cushing (1937-2002) involved, enthusiastically replied with the proposal of organising a volume of invited papers dedicated to the work of John Bell on FQM. So, with the help of the editor of the journal \emph{Foundations of Physics}, Alwyn Van der Merwe (1927-), they started inviting some of the most prominent physicists and philosophers with the aim of finally giving the long-overdue credit to Bell's work on quantum foundations. Bell was of course unaware of this project, orchestrated behind his back to positively surprise him. Writing the papers and putting together the volume,\footnote{The papers were actually published in four consecutive dedicated issues in \emph{Foundations of Physics} \cite{bellmemory}} however, took two years and the first dedicated issue appeared on October 1, 1990. Most sadly, Bell died unexpectedly on that very day, without knowing of the honour that he was about to receive \cite{bertlmanninterview}.

\subsection{Foundations of Quantum mechanics “come out of the closet”, also in Vienna}\label{sec:sub sec}

Ironically, the interest on FQM took off short after Bell's death in the 1990s, with the advent of quantum information theory and applications to quantum communication (see, e.g., \cite{gilder, freire2014}); also in Vienna this was the case.
 
 In April 1990, both Zeilinger and Berltmann helped organise the “Wolfang Pauli Symposium", held at the University of Vienna to commemorate 90 years from Pauli's birth. The keynote speaker of the conference was Weisskopf, who, invited by Walter Thirring, gave a talk dealing not  only with physics but also with his experience as an assistant of Pauli, and the intellectual atmosphere in Vienna in his youth (for instance, Sigumund Freud was a family friend). This turned out to be the last Visit of Weisskopf to Vienna.\footnote{Personal communications from Zeilinger and Bertlmann, respectively on September 17 and 20, 2021.} 
However, curiously enough, Bertlmann and Zeilinger -- despite that symposium, although they have exactly the same age, both studied at the University of Vienna, and both had a common interest in FQM (at least throughout the whole decade of 1980s)-- did not get to know each other until 1991. The occasion for their acquaintance was provided by the conference in the memory of John Bell, “Bell's theorem and the foundations of modern physics", which was organised by Selleri and collaborators on October 7-10, in Cesena (Italy). Among the speakers, which also included prominent philosophers --such as Max Jammer (1915-2010) and Paul Feyerabend-- were the two Austrian physicists who finally met and immediately started planning to organise initiatives which could bring FQM “out of the closet" in Austria too. Indeed, Bertlmann recalls: “There I met [Zeilinger], we found common interests and we thought, let’s begin to do something. And we had this idea: let’s educate young people and with this young people we can do something. Then we thought the best could be to organise a joint seminar" \cite{bertlmanninterview}. And indeed, in 1994, they managed to establish the seminar “Foundations of Quantum Mechanics" as an official  course between the University of Vienna,\footnote{Here the course was activated thanks to the support of Pietschmann, in his role of head of the Physics Institute.} where Bertlmann was associate professor, and the University of Innsbruck, where Zeilinger had just become full professor. The seminar, which had a “quite informal, familiar character" \cite{bertlmann2017}, 
 turned out to be a turning point in the acceptance of FQM at the University of Vienna, for it sensitised many students towards foundational problems that were never discussed in official courses. Indeed, short afterwards,  the first Diploma theses (the equivalent of a Master level) on FQM  started appearing: \v Caslav Brukner (1967-) --who was to become full professor of “Quantum Information Theory and Foundations of Quantum Physics" at the University of Vienna-- was the first to complete his thesis,  “Information in Individual Quantum Systems”, at the University of Vienna under Zeilinger (who was still in Innsbruck) and Pietschmann, in 1995.\footnote{Also Brukner attended the seminar on FQM, but he was not attracted into the field by it. During his undergraduate studies in physics at the University of Belgrade (Serbia) he had already being interest in entanglement thanks to his professors Fedor Herbut (1932-) and Milan Vujicic. Both had met Zeilinger at the conference for the fiftieth anniversary of EPR in Finland in 1985, and recommended Brukner to move to Vienna for working with Zeilinger.} Whereas, in 1997, Dominique Groß with the thesis “\emph{Die Nichtlokalit\"at in der Quantenmechanik}" (“Nonlocality in Quantum Mechanics"), supervised by Bertlmann,  was the first student to have directly been influenced by the quantum foundation seminar.\footnote{Meanwhile, at the University of Innsbruck, Zeilinger was supervising the first Ph.D. thesis, completed by Thomas Herzog in 1995, during which he had worked at the experimental realization of interaction-free measurements and of a quantum eraser.} Since 1996, Bertlmann also modernised the way quantum mechanics was taught in basic undergraduate courses, discussing “what a wave-function means, what is a wave-function collapse" and introducing the density matrix formalism and Bell's inequalities \cite{bertlmanninterview}. 

As for research on FQM in the 1990s, in Sect. \ref{rauch zeilinger} we have already outlined some of the results achieved by Zeilinger and his group in Innsbruck. In Vienna, at the \emph{Atominstitut} and at the Institute for Theoretical Physics of the TU, a few more people had started doing research on quantum foundations such as Johannes Summhammer, G\"unter Krenn (also in collaboration with Zeilinger) and Karl Svozil (some of their first works are \cite{tu1995, tu1996a, tu1996b}). In particular, Summhammer published one of the first partial reconstructions of quantum, deriving Malus' law from first principles  \cite{summhammer}, whereas Svozil carried out extensive research on quantum contextuality. \footnote{We are thankful to \v Caslav Brukner for pointing out this to us in personal communications on October 19, 2020 and September 16, 2021.}
 Bertlmann, too, finally entered active research in FQM in the mid-1990s, together with Walter Grimus (1953-), also a professor of theoretical particle physics at the University of Vienna. They investigated the possibility of performing Bell's experiments with massive particles, such as B-meson pairs, that are naturally produced in entangled states in accellerators, due to conservation laws (their collaboration lasted about a decade, and was to include several of their students and collaborators; the first paper is \cite{grimus}). 

When Zeilinger came back to Vienna in 1999 (this time to stay) together with Bertlmann they organised a big conference on FQM, “Quantum [Un]Speakables”. The conference took place at the University of Vienna on November 10-14, 2000, to commemorate the tenth anniversary of Bell's death and gathered the most distinguished scholars working in FQM and quantum information (and a few from particle physics). Symbolically taking place at the turn of the new century, this conference somehow marks the acceptance of FQM into the domain of physics also in Vienna, as the following anecdote encapsulates. Among the audience of the conference was Walter Thirring, the physicist who more than anyone else personified the renaissance of modern physics after World War II in Vienna (see Sect. \ref{wthirring}). He had been a leading figure in CERN, and always valued Bell very much as a particle physicist, but he realised that he had been, like most of his colleagues, quite blind towards Bell's results in quantum foundations. Yet, at the conference, he pronounced the following wistful words: “I have to apologize to John Bell, that I recognized the significance of Bell’s theorem only so late.”\footnote{Email from Bertlmann to one of the authors (F.D.S.) on October 26, 2016. Actually, Thirring went so far as to engage himself in some research on entanglement and Bell's inequalities in the following years (the first paper, with Bertlmann, is \cite{thirring2002}).}

\section{FQM in today's Vienna}\label{sec:today}

Since the 2000s, Vienna saw an exceptional growth of the activities of quantum information science and quantum technology, all branches of physics stemming more or less directly from FQM.\footnote{We will here shamelessly omit all of these more recent developments, for it would go beyond the scope of the present research whose aim was to reconstruct the pre-history of the “Zeilinger era" in Vienna. We refer the reader to \cite{zeilinger2017} for an overview.} Zeilinger, who is now Emeritus Professor at University of Vienna and the President of the Austrian Academy of Science, is today recognised among the world's leading physicists for his work in the field of FQM.\footnote{Zeilinger won several international awards, such as the prestigious Wolf Prize in Physics, which he shared with John Clauser and Alain Aspect, “for their fundamental conceptual and experimental contributions to the foundations of quantum physics, specifically an increasingly sophisticated series of tests of Bell's inequalities, or extensions thereof, using entangled quantum states." (See the website of the Wolf Fund: \href{https://wolffund.org.il/2018/12/11/anton-zeilinger/}{https://wolffund.org.il/2018/12/11/anton-zeilinger/})} Moreover,
As of the date of publication the Faculty of Physics of the University of Vienna has 5 full professors (Markus Arndt, Markus Aspelmeyer, \v Caslav Brukner, Philip Walther, and Norbert Schuch) --the first 4 of whom were either PhD students or Postdoctoral researchers under Zeilinger-- and 3 assistant professors (Borivoje Dakic, Thomas Juffmann, and Nikolai Kiesel) in the department of “Quantum Optics, Quantum Nanophysics and Quantum Information", most of whom deal with either theoretical or experimental problems of FQM; the same is true at the IQOQI-Vienna which encompasses another 6 research groups, not counting the overlaps with the Faculty (Marcus Huber, Markus M{\"u}ller, Miguel Navascués, Rupert Ursin, the Young Independent Reasearch Group, and Anton Zeilinger). These two institutions together amount to the impressive total number of roughly 150 researchers (including PhD students) working in this field. 

\section{Conclusion}\label{sec:conc}
In the present paper, we have shown that today's outstanding research landscape on FQM in Vienna did not come out of the blue by means of a single physicist (or a small group thereof), but rather developed through a long tradition of interest toward philosophical and foundational problems. Starting with the late nineteenth and early twentieth century, we have shown how the debate between Boltzmann and Mach as well as the subsequent formation of both the Exner Circle and the Vienna Circle prepared the philosophical and scientific initial conditions for foundational research in Vienna to prosper. We have discussed the impact of the advent of Nazism and the subsequent Second World War on science in Vienna and the reconstruction of physics thereafter. In particular, we have elucidated the roles of Walter Thirring, who helped bring back Vienna on the international landscape of physics, and of the PNA, founded by Herbert Pietschmann and joined by Roman U. Sexl, which reintroduced the discourse between philosophy and physics to the Faculty of Physics. Finally, we have shown how modern developments unfolded and how Vienna became today's  center of foundations of physics, by highlighting the roles of, among others, Helmuth Rauch, Anton Zeilinger, Reinhold Bertlmann as well as external influences from, e.g., Franco Selleri, Jean-Pierre Vigier, Karl Popper and John Bell.

In conclusion, it was the philosophical inclination of generations of Viennese physicists as well as the initiatives that they undertook, although those often remained  hidden from the public appearance, that arguably played a crucial role in opening room for the establishment of the present-day thriving environment of foundational research in Vienna.

 
\acknowledgements
We would like to thank Maria Irakleidou who helped carrying out this research and prepare some of the interviews. We gratefully acknowledge Herbert Pietschmann, Reinhold Bertlmann, Peter Aichelburg, Hannelore Sexl and Helmut Rauch for kindly accepting to being interviewed and for their essential support. Furthermore, with this paper we wish to honor the memory and the legacy of Helmut Rauch who passed away in September 2019 and could unfortunately not see the result of this historiographical investigation. We also would like to thank Franz Sachslehner, \v Caslav Brukner, Anton Zeilinger, Olival Freire, Jr., Mateus Ara\'ujo, Marcus Huber, Matteo Collodel and Angelo Baracca for interesting discussions and/or for providing us material from which this paper benefited. Furthermore, we are indebted to Peter Graf, Luigi Romano, and Nicole Sager for granting us access, respectively, to the archives of the Austrian Central Library for Physics (University of Vienna), the papers of Franco Selleri (Bari) and the Karl Popper Sammlung (AAU, Klagenfurt, Austria).

F.D.S. also acknowledges the financial support through a DOC Fellowship of the Austrian Academy of
Sciences (\"{O}AW).
E.S. acknowledges the support from the Austrian Science Fund (FWF) through the START project Y879-N27 and the ESQ Discovery grant ``\emph{Emergence of physical laws: From mathematical foundations to applications in many body physics}''.


\bibliographystyle{apalike}

\newpage
\hypertarget{sec:appendix}
\onecolumngrid
\appendix

\renewcommand{\thesubsection}{\thesection.\arabic{subsection}}
\renewcommand{\thesubsubsection}{\thesubsection.\arabic{subsubsection}}




\begin{small}

\end{small}

\end{document}